\documentclass[12pt]{elsart}
\usepackage{graphicx}
\usepackage{amssymb}
\usepackage{amsmath}

\journal{}

\DeclareMathOperator{\tr}{tr}
\DeclareMathOperator{\Tr}{Tr}
\newcounter{addeq}

\begin{document}
%
\begin{frontmatter}
\title{Massless excitations at $\theta =\pi$ in
the $CP^{N-1}$ model with large values of $N$}
\author[A]{A.\,M.\,M.\,Pruisken\corauthref{lc1}}
\corauth[lc1]{Corresponding author. Fax: +31 20 525 57 88
(A.\,M.\,M.\,Pruisken)} \ead{pruisken@science.uva.nl}
\author[A,I,I1]{I.\,S.\,Burmistrov}
\ead{burmi@itp.ac.ru }
\author{and\,\,\,}
\author[D]{R.\,Shankar}
\ead{shankar@imsc.res.in}
\address[A]{Institute for Theoretical Physics, University of
Amsterdam, Valckenierstraat 65, 1018 XE Amsterdam, The
Netherlands}
\address[I]{L.D. Landau Institute for Theoretical
Physics RAS, 117940 Moscow, Russia}
\address[I1]{Department of Theoretical Physics, Moscow Institute of
Physics and Technology, 141700 Moscow, Russia}
\address[D]{The Institute for Mathematical Sciences, CIT Campus, Chennai 600 113,
India}

\begin{abstract}
We study the instanton vacuum of the $CP^{N-1}$ model with large
values of $N$ in $1+1$ space-time dimensions. Unlike the
longstanding claims which state that the theory always has  a mass
gap, we for the first time establish a complete {\em critical}
theory for the transition at $\theta = \pi$ obtained from a
mapping onto the low temperature phase of the $1D$ Ising model. We
derive a simple effective field theory in terms of $1D$ massless
chiral fermions. Our results include, besides a diverging
correlation length with an exponent $\nu=1/2$, exact expressions
for the $\beta$ functions. These expressions unequivocally
demonstrate that the large $N$ expansion with varying $\theta$
displays all the fundamental features of the quantum Hall effect.
\end{abstract}


\end{frontmatter}


\section{Introduction}

The quantum Hall effect~{\cite{QHE}} has remained one of the most
important and outstanding laboratory systems where the instanton
angle $\theta$ in asymptotically free field theory can be explored
and investigated in great detail. As is well known, these
topological ideas~{\cite{TopIdeas}} originally arose in the
attempt to reconcile von Klitzing's experimental
discovery~{\cite{Klitzing}} with the scaling theory of Anderson
localization~{\cite{AndersonLoc}}. From the experimental side,
this theory set the stage for the observation of {\em quantum
criticality} of the quantum Hall plateau {\em
transitions}~{\cite{Pruisken4}} and this subject subsequently
became an important research objective both in the
laboratory~{\cite{Experiments}} as well as on the
computer~{\cite{Computer}}. From the theoretical side however,
major gaps remained in our understanding of the conceptual
structure of the problem. For example, it has only recently been
demonstrated that the topological concept of an instanton vacuum
quite generally displays {\em robust} topological quantum numbers
that explain the precision and stability of the quantum Hall
plateaus~{\cite{Robust}}. This fundamental strong coupling feature
of the theory had remained concealed for many years. The reason
being that the very general aspect of the $\theta$ vacuum concept,
the existence of {\em massless chiral edge excitations}, had
historically not been recognized.

The novel insight of massless edge
excitations~{\cite{MasslessNew}} has important consequences for
longstanding controversies such as the {\em quantization} of
topological charge, the existence of {\em discrete} topological
sectors in the theory as well as the meaning of {\em instantons}
and {\em instanton gases}~{\cite{Rev}}. A detailed understanding
of these issues has shown, for example, that the instanton vacuum
of the Grassmannian $U(m+n)/U(m) \times U(n)$ non-linear $\sigma$
model in two dimensions quite {\em generally} displays the
fundamental features of the quantum Hall effect and not merely in
the limit of zero number of field components $m=n=0$ ({\em replica
limit}) alone. These fundamental or {\em super universal} features
include

\begin{enumerate}

\item[(i)] {\em massless edge excitations} that are otherwise well
known to exist in quantum Hall systems;

\item[(ii)] {\em robust} quantization of the Hall conductance;

\item[(iii)]{\em gapless bulk excitations} at $\theta=\pi$ that generally
facilitate a {\em transition} to take place between adjacent
quantum Hall plateaus.

\end{enumerate}

This concept of {\em super universality} has in many ways been
foreshadowed by the original papers on the subject. It however,
clearly invalidates the many conflicting expectations and ideas on
the $\theta$ parameter and in particular the historical
papers~{\cite{LargeNexp,Affleck}} on the large $N$ expansion of
the $CP^{N-1}$ model~{\cite{CPN-1}} which is a special case of the
Grassmannian theory obtained by putting $m=1$ and $n=N$. We can
now say that these historical papers have merely promoted the
wrong physical ideas in the literature and served incorrect
mathematical objectives.

In this paper we address a particularly delicate and fundamental
issue, item (iii) above, about which there has been a great deal
of confusion over many years. The root of this confusion are long
standing claims which say that within the large $N$ expansion of
the $CP^{N-1}$ model ``the {\em mass gap} at $\theta=\pi$ remains
{\em finite}''~\cite{Affleck85},
 ``{\em no} critical exponents can be {\em defined}''~\cite{Affleck88}
etc. Upon assuming that these claims are true, the statement of
{\em super universality} of the quantum Hall effect would
obviously be incorrect and the theory would also not make much
sense. For example, the fundamental lesson that quantum Hall
physics has taught us is that one generally cannot disentwine the
problem of {\em quantum criticality} at $\theta=\pi$ from the {\em
existence} of the quantum Hall plateaus. This simply means that
one cannot make any random statements about the {\em transition}
at $\theta=\pi$ without having a detailed understanding of the
{\em quantization phenomenon} itself. This lack of knowledge is
the one of the reasons why it is always assumed incorrectly in the
literature that the physics of the quantum Hall effect is merely a
feature of the theory in the replica limit
alone~{\cite{Affleck85,Affleck88}} or, for that matter, the super
symmetric extensions of the disordered free electron
gas~{\cite{Efetov}}.

It has now been discovered~{\cite{Robust}} that the large $N$
expansion of the $CP^{N-1}$ model displays all the {\em super
universal} features of the quantum Hall effect as listed under (i)
- (iii) above, and also provides a lucid and exactly solvable
example of the quantum Hall {\em plateau transition}. For example,
for the first time explicit {\em finite size scaling} results for
the physical observables (``conductances'') have been obtained and
the expressions are found to be very similar to those observed
experimentally. Advances such as these are extremely important,
especially since we are dealing with a theory where the
information that can be extracted is very limited otherwise.

Even though the recent results~{\cite{Robust}} on finite size
scaling have clearly demonstrated that a divergent correlation
length exists at $\theta=\pi$ with an exponent equal to $1/2$,
they do not elucidate the {\em nature} of the massless
excitations. For this purpose we develop, in this paper, a
detailed {\em critical theory} of the large $N$ expansion,
obtained by identifying the critical operators and their
correlation functions. By presenting a complete theory that has
freed itself from any historical controversies or preconceived
mathematical biases, the authors essentially reestablish the
original ideas on the subject, especially where it says that {\em
gapless} excitations at $\theta=\pi$ are a generic topological
feature of the instanton vacuum concept with the replica method
only playing a role of secondary importance.~\cite{QHE,TopIdeas}

We start out, in Section~\ref{SecStart}, from the Coulomb gas
representation of the large $N$ theory which corresponds to the
geometry of a finite cylinder with radius $\beta/2\pi$ and length
$L$. The appropriate objects to consider are the local charges of
the Coulomb gas that are controlled by a fugacity $\sigma/\beta$
which is exponentially small in the linear dimension $\beta$. In
the language of the $U(1)$ gauge theory, these correspond to
Polyakov lines which wind around the cylinder. By considering the
limit of {\em infinite} cylinders (i.e. $L \to \infty$ and finite
$\beta$) we recognize these local charges in terms of the bosonic
``quarks" and ``anti-quarks" that appear in Coleman's argument for
periodicity in $\theta$~{\cite{Coleman}}. These particles have a
vanishing string tension as $\theta$ passes through $\pi$ at which
point it becomes energetically favorable for the system to
materialize a quark anti-quark pair that moves in opposite
directions toward ``edges" at spatial infinity. We identify the
Coulomb charges/Coleman's quarks/Polyakov lines in terms of {\em
critical operators} of a one dimensional critical theory. Based on
an explicit knowledge of all multi-point correlation functions we
are able to establish a one-to-one correspondence
(Section~\ref{corr}) between the series expansion of the Coulomb
gas in powers of the fugacity $\sigma/\beta$ on the one hand, and
the low temperature series expansion of the $1D$ Ising model on
the other. This correspondence immediately suggests that the
transition at $\theta=\pi$ can be mapped onto none other than the
$1D$ Ising model at low temperatures. This mapping is extremely
helpful since the Ising model, as is well known, is a prototypical
example of a {\em first order} phase transition with a {\em
divergent} correlation length $\xi$~{\cite{Baxter}}.

The remaining of this paper is largely devoted to the details of
the mapping of the large $N$ expansion onto exactly solvable
models in one dimension. We shall benefit in particular from our
introduction of a simple effective field theory in terms of {\em
massless chiral fermions} that elegantly displays the complete
operator structure of the theory as well as an underlying
orthogonal symmetry (Section~\ref{chiralF}). This theory has
previously emerged as the theory of {\em massless chiral edge
excitations} in quantum Hall systems~\cite{MasslessNew}.

Having identified a complete critical theory for the transition at
$\theta=\pi$ we next wish to use our results in order to obtain
exact expressions for the physical observables ($\sigma_{xx}$ and
$\sigma_{xy}$) which define the scaling behavior of the theory in
two dimensions with varying linear dimensions $\beta \approx L$.
By the phrase ``exact" we mean that our mapping procedure
facilitates a resummation to all orders in the fugacity
$\sigma/\beta$ of the Coulomb gas. For this purpose we investigate
finite size Ising spin chains and chiral fermions with a linear
dimension $L$. In Section~\ref{finite} we obtain expressions for
$\sigma_{xx}$ and $\sigma_{xy}$ in terms of Ising model and chiral
fermion correlations which are some of the most important results
of this paper. In Section~\ref{RGLN}, finally, we study the
consequences of our results in terms of the renormalization group.
We end this paper with a conclusion in Section~\ref{Conclusion}.
\section{$CP^{N-1}$ model with large values of $N$\label{SecStart}}
The action of the $CP^{N-1}$ model on a finite cylinder is
\begin{equation}
\label{cpnac} S=\int_{-L/2}^{L/2}dx\int_0^\beta d\tau~\left(
\frac{1}{g}\sum_{\alpha=1}^N\vert(\partial_\mu-iA_\mu)
z_\alpha\vert^2 +i\frac{\theta}{2\pi}\epsilon_{\mu\nu}\partial_\mu
A_\nu\right)
\end{equation}
where $\sum_{\alpha=1}^Nz^*_\alpha z_\alpha=1$. We define the
theory with the fields satisfying periodic boundary conditions in
the $\tau$ direction and {\em free boundary conditions} at $x=\pm
L/2$.

To motivate this choice of boundary conditions, we note that the
theory on a finite cylinder can be thought of in different
physical situations. Firstly, in the context of the low energy
dynamics of disordered electron gas in strong magnetic fields with
$\theta/2\pi$ denoting the mean field value of the Hall
conductance or, equivalently, the filling fraction of the Landau
bands. It also describes the long wavelength behaviour of a
dimerised $SU(N)$ quantum spin chain at temperature $\beta^{-1}$
where $\theta/2\pi$ is related to the degree of dimerisation.
Finally, it can be thought of as the finite temperature theory of
$N$ charged relativistic scalar particles in one spatial
dimension, strongly interacting with $U(1)$ gauge fields and in
the presence of a background electric field proportional to
$\theta/2\pi$.

In the case of the electronic system, it is well known that edge
currents exist and that they are crucial to the phenomenon of the
quantum Hall effect. The spin chain has dangling spins at the
edges which are the low energy degrees of freedom in the strongly
dimerised limit. In the context of scalar electrodynamics,
Coleman's picture \cite{Coleman} leads us to expect charged
degrees of freedom (``quarks'' and ``anti quarks'') at the edges.
Thus in all the three cases, we have reason to expect that the
fluctuations of the fields at the boundary play an important role
in the physics.

In what follows we shall hardly distinguish between these three
different physical interpretations of the theory and frequently
make use of any one of them, wherever it is convenient.

\subsection{Sine-Gordon model \label{SineGordon}}
In previous papers we have introduced $1D$ sine-Gordon model or
Coulomb gas representation for the large $N$ expansion of the
$CP^{N-1}$ model that effectively describes the $\theta$
dependence for a finite two dimensional cylindrical geometry with
edges~\cite{Robust}. Even though ideas very similar to ours have
been proposed a long time ago~\cite{Affleck}, the most fundamental
aspects of the problem have nevertheless been overlooked and the
exact meaning of the Coulomb gas representation has therefore not
been understood until recently. We shall proceed by recalling the
main features of the theory as it now stands.

The $z_\alpha$ fields can be integrated out in the standard large
$N$ saddle point approximation and the partition function is
written as
\begin{equation}
 Z=
 \int \mathcal{D}[A_\mu] \exp \Biggl \{- \int_0^\beta d\tau
 \int_{-\frac{L}{2}}^{\frac{L}{2}} dx
 \Bigl [ \frac{1}{2g}F_{\mu\nu}F_{\mu\nu}+
 \frac{2 \sigma}{\beta^2} (1- \cos \beta \tilde A_0) +i\frac{\theta}{4\pi}
 \epsilon_{\mu\nu}F_{\mu\nu}\Bigr ] \Biggr \}.
 \label{Z2d}
\end{equation}
Here, the coupling constant $g$ and fugacity $\sigma$ are
expressed in terms of the mass $M$ of the large $N$ expansion
according to
\begin{equation}
 g = \frac{24\pi M^2}{N},\qquad
 \sigma = N \sqrt{\frac{\beta M}{2\pi}} e^{-\beta M}
 \ll 1 .
\end{equation}
The $\tilde A_0(x)$ is the zero frequency component of
$A_0(x,\tau)$ defined as
\begin{equation}
\label{atildef} \tilde A_0(x)\equiv \frac{1}{\beta}\int_0^\beta
d\tau~A_0(x,\tau).
\end{equation}
It follows that $\exp(i\beta \tilde A_0(x))$ (the Polyakov line)
is a gauge invariant quantity and hence so is ${\rm
cos}(\beta\tilde A_0(x))$. Eq. (\ref{Z2d}) shows that the
different frequency components of $A_\mu$ are decoupled and only
the zero frequency sector depends on $\sigma$ and $\theta$.
Therefore, hereafter we concentrate solely on the zero frequency
sector. This sector is independent of $A_x$ and depends only on
$\tilde A_0$. We will also change notation and henceforth refer to
$\tilde A_0(x)$ as $A_0(x)$.

As in previous work, we separate the edge and bulk degrees of
freedom. Since the action is periodic under $A_0\to
A_0+\frac{2\pi}{\beta}$, we introduce the following resolution of
identity in the path integral,
\begin{equation}
\label{resid} 1=\int_{-\pi/\beta}^{\pi/\beta} da^l_0da^r_0
~\sum_{m_l,m_r}e^{im_l\beta(A_0(L/2)-a^l_0)}
~e^{im_r\beta(A_0(-L/2)-a^r_0)}.
\end{equation}
The partition function then gets written as,
\begin{eqnarray}
\nonumber
 Z&=&\int_{-\pi/\beta}^{\pi/\beta} da^l_0da^r_0~Z[a_0^l,a_0^r],\\
 Z[a_0^l,a_0^r]&=& \sum_{m_l,m_r} Z(m_l,m_r) \exp (- i \beta a_0^l m_l +
 i \beta a_0^r m_r ) \label{coulgas}
\end{eqnarray}
with the following meaning of the symbols. The $Z(m_l,m_r)$ is
defined as an (unconstrained) integral over static scalar
potential field $A_0 (x)$ according to
\begin{equation}
 Z(m_l ,m_r ) =
 \int \mathcal{D}[A_0] \exp \Biggl \{- \beta \int_{-\frac{L}{2}}^{\frac{L}{2}} dx
 \Bigl [ \frac{1}{g} ( \partial_x A_0)^2 +
 \frac{2 \sigma}{\beta^2} (1- \cos \beta A_0) +i
 A_0 \rho_m \Bigr ] \Biggr \}. \label{Zmlmr}
\end{equation}
Here, $\rho_m$ represents the {\em integral} charges $m_l , m_r$
as well as {\em external fractional} charges $\theta /2\pi$
located at the opposite edges $x = \pm L/2$ of a system of linear
spatial dimension $L$. It is given by
\begin{equation}
 \rho_m (x) =\left( m_l+\frac{\theta}{2\pi}\right)\delta( x + L/2)
 +\left( m_r-\frac{\theta}{2\pi}\right)\delta( x - L/2).
\end{equation}
One of the most significant quantities are phase factors $e^{- i
\beta a_0^l m_l}$ and $ e^{ i \beta a_0^r m_r}$ in Eq.
\eqref{coulgas} which can be interpreted in terms of the {\em
fluctuations} in the topological charge of the theory about its
integral values. In terms of the original $CP^{N-1}$ variables,
these phase factors are the Berry phases that appear in the time
evolution of $s= |m_{l,r}|/2$ $SU(N)$ {\em spin} degrees of
freedom located at the edge ($x=\mp L/2$)
~\cite{PruiskenShankarSurendran}. More specifically, $\exp(-i
\beta a_0^l m_l)$ and $\exp(i \beta a_0^r m_r)$ are given by the
following,
\begin{equation}
 e^{- i \beta a_0^{l} m_{l}} = \exp S_{l} [z^*,z],\qquad
 e^{i \beta a_0^{r} m_{r}} = \exp S_{r} [z^*,z].
 \label{chiral1}
\end{equation}
The action $S_{l,r}$ for spin dynamics is given in terms of the
original $CP^{N-1}$ complex vector field variables $z_\alpha$
according to
\begin{equation}
 S_{l,r} [z^*,z] = |m_{l,r}| \int_0^\beta
 d\tau ~ \left[ \mp  z^*_\alpha~ \partial_0 ~{z}_\alpha + b z^*_1 z_1
 \right].
 \label{chiral2}
\end{equation}
Here, the quantity $b$ denotes an external magnetic field which
defines the theory in the infrared.
\subsection{Coulomb gas representation}
By series expanding Eq. \eqref{coulgas} in powers of $\sigma$ one
immediately sees that the $\cos \beta A_0$ term generates local
charges with $\sigma/\beta$ playing the role of fugacity. We can
write
\begin{eqnarray}
 & & \exp\frac{2\sigma}{\beta}\int_{-L/2}^{L/2} dx \, \cos \beta A_0 =
 \sum_{n_\pm =0}^\infty \frac{ (\sigma/\beta)^{n_+ + n_-}
 }{n_+ ! n_- !}\hspace{1cm}{}
 \notag \\
 &&\hspace{2cm}\times \prod_{k=1}^{n_+} \int_{-L/2}^{L/2}
 dx_k^+\prod_{j=1}^{n_-} \int_{-L/2}^{L/2} dx_j^- \exp i \beta \int_{-L/2}^{L/2} dx\, A_0 (x
 ) \rho_n (x )
 \label{CosRep}
\end{eqnarray}
where $\rho_n$ denotes the {\em charge density} in the bulk of the
system
\begin{equation}
 \rho_n (x) = \sum_{k=1}^{n_+} \delta(x -x_i^+ ) -
 \sum_{j=1}^{n_-} \delta(x - x_j^- ) .
\end{equation}
The free field $A_0$ can now be eliminated and the final result
can be written as
\begin{equation}
 Z(m_l ,m_r ) = \sum_{n_\pm =0}^\infty
 \frac{\delta_{m_l-m_r,n_{-}-n_+}}{n_+ ! n_- !}
 { \left( \frac{\sigma}{\beta} \right)^{n_+ + n_-} }~
 \prod_{k=1}^{n_+} \prod_{j=1}^{n_-} \int_{-L/2}^{L/2} dx_k^+  \int_{-L/2}^{L/2}
 dx_j^-\, e^{- \beta
 \mathcal{H}_\textrm{coul}}\label{Zmm} .
\end{equation}
The hamiltonian
\begin{equation}
 \mathcal{H}_\textrm{coul} = -\frac{g}{8} \int_{-L/2}^{L/2}  \int_{-L/2}^{L/2}
 dxdy ~\rho_{mn}
 (x) |x-y| \rho_{mn} (y )\label{Hcoul} .
\end{equation}
describes a system of interacting charges in one dimension. The
total charge density is given by the sum of the edge parts
$\rho_m$ and bulk parts $\rho_n$
\begin{equation}
 \rho_{mn} (x) = \rho_n (x) + \rho_m (x ) .
\end{equation}
Finally, the symbol $\delta_{m_l-m_r,n_{-}-n_+}$ in Eq.
\eqref{Zmm} ensures the charge neutrality of the Coulomb gas.
\section{Quantum Hall effect \label{Sec3B}}
\subsection{Introduction}
The results of the previous Sections are unchanged when considered
as a model of the quantum Hall effect. The only difference is that
the imaginary time $\tau$ now plays the role of the $y$-coordinate
such that $L$ and $\beta$ represent the linear dimensions of the
system in the $x$ and $y$ directions respectively. The action
$S_{l,r}$ in Eq. \eqref{chiral2} now describes the {\em massless
chiral edge excitations} in the problem rather than {\em spin
dynamics}, and the quantity $b$ stands for external frequency
rather than magnetic field. Furthermore, to retain the notation of
previous work we shall from now onward use the symbol $\kappa$
rather than $g$, i.e.,
\begin{equation}
\kappa = \frac{1}{2\beta g} = \frac{N}{48\pi M^2 \beta} .
\end{equation}
The remarkable thing about the massless edge excitations is that
they are identically the same for all members of the $U(m+n)/U(m)
\times U(n)$ non-linear $\sigma$ model. This previously
unrecognized aspect of the problem is the sole reason why the
``conductances" or ``response parameters" $\sigma_{xx}$ and
$\sigma_{xy}$, to be discussed further below, are in fact the most
important physical quantities in the problem that are uniquely
defined for all values of $m$ and $n$. This is in spite of the
fact that their true significance as transport coefficients is
retained in the theory with $m=n=0$ only.

As has been explained at many place elsewhere, the correct
expressions for the ``response parameters" emerge from the
definition of the {\em effective action} for the edge field
variables $z_\alpha$ and $z_\alpha^*$ and can quite generally be
considered as a measure for the response of the interior of the
system to infinitesimal changes in the boundary conditions. For
this purpose it is convenient to employ the quantities $\beta
a_0^l$ and $\beta a_0^r$ in Eq. \eqref{coulgas}. By expanding the
free energy in small fluctuations $a_0^l$ and $a_0^r$ it can be
shown that
\begin{equation}
 \ln Z[a_0^l,a_0^r] = -F(\theta) + i
 {\langle \sigma_{xy}^\prime \rangle } \beta (a_0^l - a_0^r)
 - {\langle \sigma^\prime_{xx} \rangle} \left( \beta \frac{ a_0^l +
a_0^r}{2}\right )^2 ~+~\dots\label{conduct}
\end{equation}
where $\langle \sigma^\prime_{xx} \rangle$ and $\langle
\sigma^\prime_{xy} \rangle$ are in all respects analogous to the
dimensionless {\em longitudinal} and {\em Hall} conductance
respectively of the disordered free electron gas of size $\beta
\times L$. Similarly, the coefficients of the higher order terms
in the series of Eq. \eqref{conduct} can be interpreted in terms
of conductance {\em fluctuations} or conductance {\em
distributions}. An interesting feature of the large $N$ expansion
is that these conductance distributions can quite generally be
expressed in terms of the {\em ensemble averaged} quantities
$\langle \sigma^\prime_{xx} \rangle$ and $\langle
\sigma^\prime_{xy} \rangle$ alone. We will come back to this point
at the end of Section~\ref{plateautrans}.
\subsection{The quantum Hall state, $\theta \approx 2\pi k$}
To understand how the theory generates the physics of the quantum
Hall effect let us first evaluate the partition function of the
Coulomb gas, Eq. \eqref{Zmm}, to lowest orders in the fugacity
$\sigma/\beta$. By considering the terms with $(m_l, m_r)=(m, m)$
and $(m\pm 1, m)$ for an arbitrary integer $m$ we obtain the
following expression~\cite{Robust}
\begin{equation}
 Z[a_0^l, a_0^r] =
 \sum_{m\in\mathbb{Z}} \zeta (m) e^{-\frac{L}{4\kappa} \left
 (m+\frac{\theta}{2\pi} \right )^2 + i m \beta ( a_0^r -
 a_0^l )} \label{partition}
\end{equation}
where
\begin{equation}
\zeta (m) = 1 -  \frac{4\kappa\sigma}{\beta} \frac{e^{i \beta
a_0^l} + e^{-i \beta a_0^r }}{\frac{\theta}{\pi} +2m -1} +
\frac{4\kappa\sigma}{\beta}\frac{e^{-i \beta a_0^l} + e^{i \beta
a_0^r }}{\frac{\theta}{\pi} +2m +1}.\label{partition2}
\end{equation}
Next, we take the quantity $\theta/2\pi$, representing the {\em
filling fraction} of the Landau bands, to lie in the interval
$k-1/2 < \theta/2\pi < k+1/2$ for an arbitrary integer $k$. It is
easy to see that the partition function of Eq. \eqref{partition}
is then dominated by the single term with $m=-k$ and for large
system sizes $\beta L$ we therefore have
\begin{equation}
 Z[a_0^l, a_0^r] =
 \exp\left [-\frac{L}{4\kappa} \left
 (\frac{\theta}{2\pi}-k \right )^2 +  i k \beta ( a_0^r -
 a_0^l )\right ] + \dots \label{qHstates}
\end{equation}
All the other terms represented by $\dots$ are smaller by factors
that are exponential in $\beta$ or $\beta L$. By comparing the
result with Eq. \eqref{conduct} we now recognize the integer $k$
as the {\em robustly} quantized Hall conductance. More precisely,
we can say that for all {\em filling fractions} in the range
$k-1/2 < \theta/2\pi < k+1/2$ we have $\langle \sigma_{xy}^\prime
\rangle =k$ and $\langle \sigma_{xx}^\prime \rangle = 0$ except
for corrections that are exponentially small in the system size.
The results are therefore precisely in accordance with the
experimental observations of the quantum Hall effect.

Notice that in the language of the Coulomb gas the {\em quantum
Hall state}, labelled by the integer $k$, is synonymous for having
``quarks" and ``anti quarks" at the edges of the system such as to
maximally shield the fractional charges $\pm \theta/2\pi$.
However, besides the quantized {\em charges} $\pm k$, the
``quarks" and ``anti quarks" at the edges also carry {\em spin}
degrees of freedom. Following the discussion in Section
\ref{SineGordon} we conclude that the complete expression for the
partition function for the {\em quantum Hall state} reads, instead
of Eq. \eqref{qHstates},
\begin{equation}
 Z[a_0^l, a_0^r] \rightarrow
 e^{-F (\theta) } ~ Z_l ~ Z_r  \label{qHstates1}
\end{equation}
where
\begin{equation}
 F (\theta) = \frac{L}{4\kappa} \left
 (\frac{\theta}{2\pi}-k \right )^2,\qquad
 Z_{l,r} = \int \mathcal{D}[z^* z] ~ e^{S_{l,r}[z^*,z]} \label{qHstates2}
\end{equation}
denote the bulk free energy and the one dimensional partition
functions for the edge spins respectively. Here, the action for
the edge $S_{l,r}$ is the same as in Eq.~\eqref {chiral2} but with
the integer $k$ now replacing the spin quantum numbers $m_l$ and
$m_r$.

From the results of this Section it is clear that the
identification of Eqs \eqref{qHstates1} and \eqref{qHstates2} with
the quantum Hall state is solely based on the edge parts of the
theory $Z_{l,r}$ that describe the well known edge currents in the
problem. These subtleties of the edges have historically not been
recognized, however.
\subsection{Plateau transitions,
$\theta \approx 2\pi (k+1/2)$ \label{plateautrans}}
\begin{figure}[tbp]
\begin{center}
\includegraphics[width=120mm]{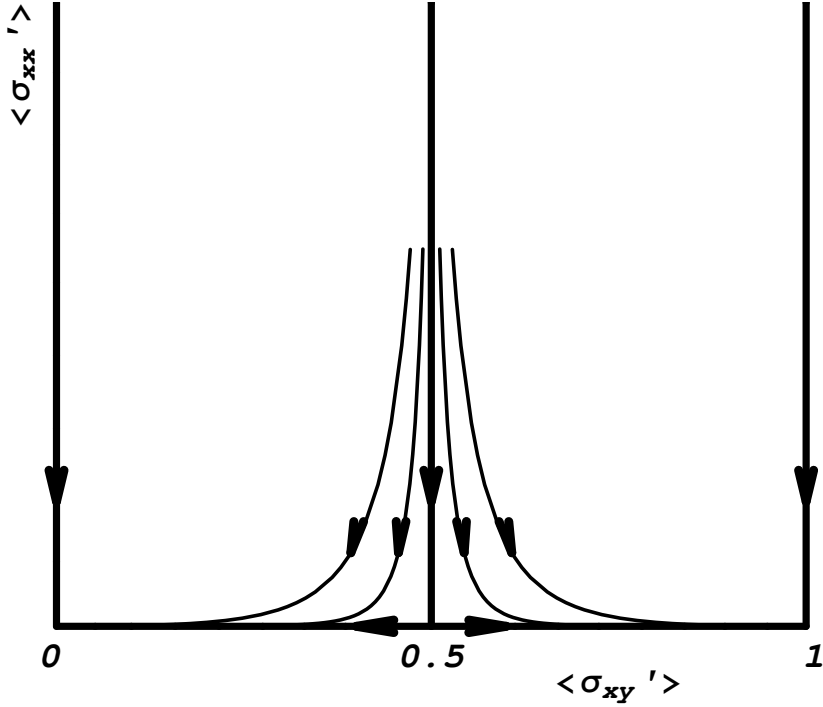}
\end{center}
\caption{Renormalization group flow in the $\langle
\sigma^\prime_{xx}\rangle$ and $\langle \sigma^\prime_{xy}\rangle$
conductance plane according to Eqs \eqref{kubo1} -
\eqref{XetaCG2}, see text.}\label{RGFig1s}
\end{figure}

A {\em transition} takes place between the adjacent quantum Hall
states, labelled by the integers $k$ and $k+1$, at the exact
values $\theta /2\pi = (k + 1/2)$ which in the language of the
electron gas corresponds to half-integer filling fractions. Notice
that in the limit where $\beta L$ tends to infinity this {\em
plateau} transition is infinitely sharp. Similarly, from Eq
\eqref{qHstates} we conclude that, in the same limit, the free
energy of the bulk $F(\theta)$ develops a cusp at $\theta = (2k +
1)\pi$ according to
\begin{equation}
 F (\theta) \simeq -\frac{|2k+1 -\theta/\pi|}{8\kappa} L  \label{FlargeN}
\end{equation}
indicating that the transition is a {\em first order} one. Next,
to develop a better understanding of the {\em nature} of the
plateau transitions we evaluate the expression for the partition
function Eqs ~\eqref{partition} and \eqref{partition2} for
$\theta$ close to an odd multiple of $\pi$. Taking $k=0$ for
simplicity then the sum in Eq. \eqref{partition} is dominated by
the terms with $m=0,-1$ and the result can be written as
\begin{equation}
 Z[ q, a_0] = e^{-F (\theta)}
 \left ( \mathcal{P}_0 e^{0 i q}
 + \mathcal{P}_\pi
 e^{\pi i q} \cos \beta a_0 +
  \mathcal{P}_{2\pi}
 e^{2 \pi i q}
 \right ) .\label{free}
\end{equation}
We have introduced the symbols
\begin{equation}
 q = \frac{\beta}{2\pi}(a_0^l - a_0^r) ,\qquad a_0 = \frac{1}{2} (a_0^l + a_0^r).
\end{equation}
The quantities $\mathcal{P}$ sum up to unity, $\mathcal{P}_0 +
\mathcal{P}_\pi + \mathcal{P}_{2\pi} =1$, and can be expressed as
follows
\begin{eqnarray}
 \mathcal{P}_0 &=& 1-{\langle \sigma_{xy}^\prime \rangle} - \langle
 \sigma_{xx}^\prime \rangle  \\
 \mathcal{P}_\pi &=& 2 \langle \sigma_{xx}^\prime \rangle \\
 \mathcal{P}_{2\pi} &=& {\langle \sigma_{xy}^\prime \rangle}- \langle
 \sigma_{xx}^\prime \rangle .\label{pees}
\end{eqnarray}
The following expressions for the ``ensemble averaged"
conductances $\langle \sigma_{xx}^\prime \rangle$, $\langle
\sigma_{xy}^\prime \rangle$ have been obtained~\cite{Robust}
\begin{eqnarray}
 \langle \sigma_{xx}^\prime \rangle &=& \frac{\eta }{e^{X}+e^{-X} +
 2  \eta} \label{kubo1} \\
 \langle \sigma_{xy}^\prime \rangle &=& \frac{e^{-X}+  \eta}{e^{X}+e^{-X} +
 2 \eta} .\label{kubo2}
\end{eqnarray}
The two different scaling variables $X$ and $\eta$ are given as
\begin{eqnarray}
 X &=& \frac{L}{8\kappa} \left (1-\frac{\theta}{\pi}\right ) \label{XetaCG1}\\
 \eta &=& \frac{L}{\beta} \sigma
 \frac{\sinh X}{X}.
 \label{XetaCG2}
\end{eqnarray}
These explicit expressions which are defined for a system with
linear dimensions $\beta$ and $L$ play a central role in the
remainder of this paper. The most important features of these
finite size scaling results are the symmetry about $\theta=\pi$
(``particle-hole" symmetry) and the fact that they display all the
characteristics of a {\em continuous} transition with a {\em
diverging} correlation length $\xi$. To see this we put $\beta=L$
which is the most natural geometry to consider (see, however, the
discussion at the end of this Section). The scaling variable $X$
in Eq. \eqref{XetaCG1} can then be written in the following manner
\begin{equation}
 X = \pm \frac{L^2}{\xi^2}, \qquad \xi \propto
 \left |1-\frac{\theta}{\pi}\right |^{-1/2}. \label{xi}
\end{equation}
The results clearly show that the theory at $\theta=\pi$ must have
{\em gapless} excitations. Moreover, by expressing
Eqs~\eqref{kubo1} - \eqref{kubo2} in differential form we obtain
the following general results for the $\beta$ functions
\begin{eqnarray}
 \frac{d \langle \sigma_{xx}^\prime \rangle }{d \ln L} &=&
 \beta_{xx} \left( \langle \sigma_{xx}^\prime \rangle , \langle
 \sigma_{xy}^\prime
 \rangle \right) \label{RGeqns1}\\
 \frac{d \langle \sigma_{xy}^\prime \rangle }{d \ln L} &=&
 \beta_{xy} \left( \langle \sigma_{xx}^\prime \rangle , \langle \sigma_{xy}^\prime
 \rangle \right) .\label{RGeqns2}
\end{eqnarray}
The renormalization group flow lines in the $\langle
\sigma_{xx}^\prime \rangle$, $\langle \sigma_{xy}^\prime \rangle$
conductance plane  are illustrated in Fig.~\ref{RGFig1s}. The
general result of Eqs \eqref{RGeqns1} - \eqref{RGeqns2} has
previously been established in the weak coupling regime $\langle
\sigma_{xx}^\prime \rangle \gg 1$ based on
instantons~{\cite{Inst}}. Therefore, by combining the known weak
coupling form for the $\beta$ functions with the strong coupling
results based on Eqs~\eqref{kubo1} -\eqref{kubo2} we obtain
complete information on the general phase structure of the large
$N$ expansion. The {\em super universal} features of this theory
are the symmetry about the line $\langle \sigma_{xy}^\prime
\rangle$ equal to half integer values, the infrared {\em stable}
fixed points located at $\langle \sigma_{xy}^\prime \rangle =k$
and the {\em unstable} fixed points located at $\langle
\sigma_{xy}^\prime \rangle =k +1/2$.

Several remarks are in order. First of all, the result of Eq.
\eqref{free} indicates that the plateau transition is described in
terms of an admixture of three distinctly different {\em phases}
with a well defined {\em probability} $\mathcal{P}$ each. These
phases are labelled by an exponential factor $e^{i \theta^\prime
q}$ with $\theta^\prime$ taking on the values $0, \pi$ and $2\pi$
respectively. The quantity $\mathcal{P}_\pi$ in Eq. \eqref{free}
can be interpreted as the {\em probability} of finding the system
in the {\em dissipative} or {\em critical} phase labelled by
$\theta^\prime = \pi$. In the same way, $\mathcal{P}_{2\pi}$
denotes the {\em probability} of finding the system in the
$\theta^\prime = 2\pi$ vacuum or $k=1$ quantum Hall phase.
Similarly, $\mathcal{P}_{0}$ is the probability of finding the
system in the $\theta^\prime = 0$ vacuum or $k=0$ quantum Hall
phase. As we shall see in the analysis that follows, correlation
functions of the theory display exactly the same general structure
as that of Eqs \eqref{free} - \eqref{pees}.

Secondly, it should be mentioned that the result of Eq.
\eqref{free} determines not only the ``ensemble averaged"
conductances $\langle \sigma_{xx}^\prime \rangle$ and $\langle
\sigma_{xy}^\prime \rangle$ but also the complete statistics of
conductance {\em distributions}~{\cite{CondDistr}}. For example,
expressing Eq. \eqref{free} in the exponential form of Eq.
\eqref{conduct} then we directly see that besides the lowest order
terms in $q$ and $\beta a_0$ we actually have an infinite series
of higher order moments. These higher order moments can all be
expressed in terms of the ``ensemble averaged" quantities $\langle
\sigma_{xx}^\prime \rangle$ and $\langle \sigma_{xy}^\prime
\rangle$, however, and it suffices to limit the analysis to the
``ensemble averaged" quantities alone.

Thirdly, in spite of the rich structure that emerges, it is
important to keep in mind that we have considered the Coulomb gas
to lowest order in an expansion in powers of the fugacity
$\sigma/\beta$ only. Even though the expressions in
Eqs~\eqref{kubo1} - \eqref{kubo2} are generally well defined in
the limit $\beta \approx L \rightarrow \infty$, they nevertheless
show that the expansion in the fugacity $\sigma/\beta$ actually
diverges at $\theta=\pi$ in the limit where $L$ is taken to
infinity first. This complication in defining the thermodynamic
limit clearly indicates that the higher order terms in the
expansion are important. This will be the main subject of Section
\ref{finite} where we make use of the mapping onto the Ising model
and chiral fermion theory in order to be able to re-sum the series
in the fugacity $\sigma/\beta$ to infinite order.
%
%
\section{Correlation functions \label{corr}}
\subsection{Introduction \label{introcorr}}
To develop a theory for gapless excitations at $\theta=\pi$ we
next embark on the subject of correlation functions. We are
specifically interested in the finite temperature correlation
function of the Coulomb gas
\begin{equation}
 G_2 (x,y)=  \langle e^{i\beta A_0 (x)}e^{-i\beta A_0 (y)}\rangle =
 \left \langle \exp\left (-i\beta \int_x^y dx\, \partial_x A_0 \right )\right \rangle
\end{equation}
which corresponds to a pair of static charges or a quark
anti-quark pair at positions $x$ and $y$ respectively. In the
limit of zero fugacity ($\sigma = 0$) we immediately find from Eq.
\eqref{Zmm} (putting, for simplicity, the spin degrees of freedom
at the edges equal to zero, $q=\beta a_0 =0$)
\begin{equation}
 G_2(x,y) =
  \frac{\sum\limits_{m\in\mathbb{Z}} e^{- \frac{L\beta g}{4} (\theta/2\pi +m)^2} ~
  e^{- \frac{\beta g}{4} U_2(x,y;m)}}
  {\sum\limits_{m\in\mathbb{Z}} e^{- \frac{L\beta g}{4} (\theta/2\pi +m)^2}}
  \label{paircorr}
\end{equation}
where
\begin{equation}
  U_2(x,y;m) = (\theta/\pi +2m) (x-y) + |x-y| .\label{u2xym}
\end{equation}
We are interested from now onward in the limit $L \rightarrow
\infty$ while keeping $\beta$ fixed which, as we shall see below,
enables us to make contact with Coleman's argument for having
periodicity in $\theta$. As discussed in the previous Section,
this limit precisely corresponds to the situation where the naive
expansion in powers of the fugacity $\sigma/\beta$ gets
complicated. To deal with these complications we shall proceed and
obtain, in the remaining parts of this Section, expressions for
arbitrary multi point correlation functions of the Coulomb gas.
These expressions then serve as a starting point in
Section~\ref{IsingModel} for the mapping of the Coulomb gas
problem onto exactly solvable models in one dimension.
\subsection{Coleman's picture \label{picture}}
First, let $\theta$ approach $\pi$ from below. The sum in Eq.
\eqref{paircorr} is then dominated by the term with $m=0$ and we
can write
\begin{equation}
G_2 (x,y) = \vartheta (y-x) e^{-(1 - \theta/\pi)|y-x|/4\kappa} +
\vartheta (x-y) e^{-(1 + \theta/\pi)|y-x|/4\kappa},\,(\theta
\rightarrow \pi^-). \label{thetaminusfull}
\end{equation}
Here, $\vartheta$ denotes the Heaviside step function. On the
other hand, when $\theta$ approaches $\pi$ from above the sum is
dominated by the $m=1$ term and the result is
\begin{equation}
G_2 (x,y) = \vartheta (y-x) e^{-(3 - \theta/\pi)|y-x|/4\kappa} +
\vartheta (x-y) e^{(1 - \theta/\pi)|y-x|/4\kappa},\,(\theta
\rightarrow \pi^+). \label{thetaplusfull}
\end{equation}
In the limit of large separations one may replace the correlation
function $G_2 (x,y)$ by its long ranged part which will be denoted
by $g_2 (x,y)$
\begin{eqnarray}
 G_2 (x,y) \rightarrow g_2 (x,y) &=& \vartheta (y-x)
 e^{-(1 - \theta/\pi)|y-x|/4\kappa} ,\, (\theta
 \rightarrow \pi^-) \label{thetaminus} \\
 G_2 (x,y) \rightarrow g_2 (x,y) &=& \vartheta
 (x-y) e^{+(1 -\theta/\pi)|y-x|/4\kappa},\, (\theta \rightarrow \pi^+) .
 \label{thetaplus}
\end{eqnarray}
Eqs \eqref{thetaminus} and \eqref{thetaplus} describe the
mechanism that, following Coleman, is responsible for
adding/removing charges at the edges at infinity when $\theta$
passes through $\pi$. Eq. \eqref{thetaminus} tells us, for
example, that when $\theta$ passes through $\pi$ from {\em below},
it is energetically favorable for the system to materialize a
quark and an anti-quark that ``move" in opposite directions to the
edges such as to maximally shield the background fractional
charges $\pm \theta /2\pi$. This picture is consistent with the
result of Eq. \eqref{thetaplus} which says that when $\theta$
passes through $\pi$ from {\em above}, the pair correlation is
dominated by the $m=1$ vacuum which means that a quark and an
anti-quark are present at the edges at infinity.

Coleman's picture is strictly valid for the Coulomb gas with a
vanishing fugacity (or $\beta \rightarrow \infty$) only and the
physical processes associated with finite values of $\sigma/\beta$
or $\beta$ are by no means obvious. Nevertheless, it is extremely
important to recognize that Coleman's picture is in fact
synonymous for having {\em massless} excitations at $\theta=\pi$.
More specifically, the statements made by Eqs \eqref{thetaminus}
and \eqref{thetaplus} are the simplest possible example of a {\em
critical theory} and the scaling dimension of the {\em critical
operators} (quarks or Polyakov lines) is equal to unity. Moreover,
exponential factors such as $\exp[\pm(1-\theta/\pi)|x-y|/4\kappa]$
have a quite general significance in critical phenomena theory and
for the problem at hand they demonstrate that when $\theta$
approaches $\pi$ there is a {\em diverging} correlation length
according to $\xi = 4\kappa |1-\theta/\pi|^{-1}$.

When looked upon as a critical theory in {\em two} dimensions
rather than one, then the critical operators are nonlocal objects
and the exponential factors are more appropriately understood in
terms of an {\em area law} $\exp[\pm \beta |x-y| /\xi^2]$ where
$\xi$ is now given by Eq. ~\eqref{xi} and $\beta |x-y|$ denotes
the {\em area} enclosed by the Polyakov lines. The same area law
appears in the scaling results for the conductances, Eqs
\eqref{kubo1} - \eqref{XetaCG2}. This clearly indicates that both
Coleman's mechanism and the quantum Hall plateau transition should
generally be regarded as distinctly different consequences of the
same fundamental principle, notably the existence of {\em
massless} excitations at $\theta=\pi$.

It is important to keep in mind that the considerations of this
Section primarily apply to the Coulomb gas with zero fugacity. To
deal with the complications of the theory with finite values of
$\sigma/\beta$ we next proceed and embark on the subject of
multi-point correlation functions.
\subsection{Four point correlations ($\theta \rightarrow \pi^-$)}
Write
\begin{equation}
 G_4 (x_1 y_1 x_2 y_2 ) = \langle e^{i\beta A_0 (x_1)} e^{-i\beta A_0 (y_1)}
 e^{i\beta A_0 (x_2)} e^{-i\beta A_0 (y_2)} \rangle
\label{fourcorr0}
\end{equation}
then we immediately obtain from Eq. \eqref{Zmm}
\begin{equation}
 G_4 (x_1 y_1 x_2 y_2 ) = \frac{\sum\limits_{m\in\mathbb{Z}} e^{- \frac{L}{4\kappa}
 (\theta/2\pi +m)^2} ~
 e^{- \frac{1}{4\kappa} U_4 (x_1 y_1 x_2 y_2  ;m) }}
 {\sum\limits_{m\in\mathbb{Z}} e^{- \frac{L}{4\kappa} (\theta/2\pi +m)^2}}
 \label{fourcorr}
\end{equation}
where $U_4$ is given by
\begin{eqnarray}
  U_4(x_1 y_1 x_2 y_2 ;m) &=& (\theta/\pi +2m) ( y_1 - x_1 + y_2 - x_2 )
  + | y_1 - x_1|+| y_2 - x_2 | \nonumber \\
  &+& | y_{2} - x_1|+| x_2 - y_1 | - |y_2 - y_1| - |x_2 - x_1|
  .\label{u4xym}
\end{eqnarray}
For simplicity we consider only the case where $\theta$ approaches
$\pi$ from below. Like $G_2$ the sum in Eq. \eqref{fourcorr} is
then dominated by the $m=0$ term only. Next, assume the following
arrangement of the coordinates (cf. Ref.~{\cite{Mapping}})
\begin{equation}
 x_1 < y_1 < x_2 < y_2
  \label{llll}
\end{equation}
then Eq. \eqref{u4xym} simplifies and we can write
\begin{equation}
 U_4(x_1 y_1 x_2 y_2 ;m) = (1 - \theta/\pi) ( y_1 - x_1 + y_2 - x_2)
 .\label{u4xym1}
\end{equation}
This result is invariant under the interchange $y_1
\leftrightarrow y_2$ and  $x_1 \leftrightarrow x_2$. From Eqs
\eqref{llll} and \eqref{u4xym1} one immediately concludes that the
long ranged parts of $G_4$ can be expressed according to
\begin{eqnarray}
 G_4 (x_1 y_1 x_2 y_2 ) &\rightarrow& g_4 (x_1 y_1 x_2 y_2 )
 \nonumber \\
 &=&[ \vartheta_3 (x_1 y_1 x_2 y_2) + \vartheta_3 (x_2 y_1 x_1
 y_2)\notag \\
 &&+\vartheta_3 (x_1 y_2 x_2 y_1) + \vartheta_3 (x_2 y_2 x_1 y_1) ]
 e^{-2\omega  (y_1 - x_1 + y_2 - x_2)} \label{g4}
\end{eqnarray}
where we have introduced
\begin{equation}
 \vartheta_3 (x_1 y_1 x_2 y_2) =
 \vartheta(y_1 - x_1) \vartheta(x_2 - y_1) \vartheta(y_2 - x_2)
 \label{k}
\end{equation}
and
\begin{equation}
 \omega = \frac{1}{8\kappa}\left (1-\frac{\theta}{\pi}\right ) >0 .\label{omega}
\end{equation}
\subsection{Multi point correlations ($\theta \rightarrow \pi^-$)}
A generalization of the results of the previous Section is
straight forward. Let the $2n$-point correlation function be
denoted by
\begin{equation}
G_{2n}(x_1,y_1,\dots,x_{n},y_{n})=\left \langle \prod_{k=1}^n
e^{iA_0(x_k)}e^{-iA_0(y_k)}\right \rangle
\end{equation}
then by using Eq. \eqref{Zmlmr} we immediately find
\begin{equation}
 G_{2n}(x_1,y_1,\dots,x_{n},y_{n}) = \frac{\sum\limits_{m\in\mathbb{Z}} e^{-
 \frac{L}{4\kappa} (\theta/2\pi +m)^2} ~ e^{-\frac{1}{4\kappa}
 U_{2n}(x_1,y_1,\dots,x_{n},y_{n} ; m)}} {\sum\limits_{m\in\mathbb{Z}} e^{-
 \frac{L}{4\kappa} (\theta/2\pi +m)^2} ~} \label{multicorr}
\end{equation}
where
\begin{eqnarray}
U_{2n} = (\theta/\pi +2m ) \sum_{j=1}^{n} (x_j-y_j) &+&
\sum_{j=1}^{n} |x_j-y_j| + \sum_{j>i} (|y_j-x_{i}|+|x_j - y_{i}|
\notag \\
&-& |x_j-x_i| - |y_j-y_i| ) .
\end{eqnarray}
Consider the case where $\theta$ approaches $\pi$ from below then
the sum in Eq. \eqref{multicorr} is again dominated by the $m=0$
term only. Next, assume the following arrangement of the
coordinates
\begin{equation}
x_{1} < y_{1} < x_2 < y_2 < \dots < y_{k} < x_{{k+1}} < \dots
<y_{n}. \label{Seq}
\end{equation}
It can then easily be shown that
\begin{equation}
 U_{2n} (x_1,y_1,\dots,x_n,y_n) = \left
 (1-\frac{\theta}{\pi}\right ) \sum_{i=1}^n(y_i-x_i).
 \label{ordered}
\end{equation}
Since Eq. \eqref{ordered} is invariant under all permutations
$P(x_1 \dots x_n)$ and $P(y_1 \dots y_n)$ we finally obtain the
following total result for the long ranged parts of $G_{2n}$
\begin{equation}\label{g2nGen}
 G_{2n} \rightarrow g_{2n} = \sum_{ P(x_1\dots x_n)
 }\sum_{P( y_1\dots y_n) } \vartheta_{2n-1}
 (x_{1},y_{1},\dots,x_{n},y_{n}) \exp \left [ -2
 \omega \sum_{i=1}^n(y_i-x_i)\right ]
\end{equation}
where the first two sums are over all permutations and
\begin{equation}\label{g2nGen1}
 \vartheta_{2n-1} (x_{1},y_{1},\dots,x_{n},y_{n}) =
 \vartheta(y_{1}-x_{1}) \prod_{j=2}^n \vartheta( x_j - y_{j-1})
 \vartheta ( y_j - x_j ) .
\end{equation}
\section{ $1D$ Ising model \label{IsingModel}}
\subsection{Introduction}
Given the explicit form of the correlation functions of the
Coulomb gas with fugacity $\sigma/\beta$ equal to zero, we next
wish to exponentiate the operators $e^{\pm i \beta A_0}$ in order
to solve the problem with finite values of the fugacity. More
specifically, we are interested in the partition function of the
Coulomb gas which can be written as an operator statement
according to (see also Eq. \eqref{CosRep})
\begin{equation}
 Z_{CG} = \left \langle \exp \left [ \frac{2\sigma}{\beta} \int dx \cos \beta
 A_0 (x) \right ] \right \rangle_{\theta \rightarrow \pi^-} .\label{ZCG}
\end{equation}
By writing Eq. \eqref{ZCG} as a series expansion in powers of
$\sigma/\beta$ we can formally express the result in terms of the
multi-point correlation functions of the $m=0$ sector as
considered in the previous Section. At this stage of the analysis
it may not be entirely obvious that such a procedure provides
indeed the correct answer in the limit $L \rightarrow \infty$ and
for varying values of $\theta \approx \pi$. In the Sections below
we shall show, however, that the series expansion of Eq.
\eqref{ZCG} is in one-to-one correspondence with the well known
low temperature series expansion of the one dimensional Ising
model. Once the correspondence with the Ising model is established
we can proceed in a variety of different ways and solve the
Coulomb gas problem of Eq. \eqref{ZCG} based on the results
obtained from certain exactly solvable models.
\subsection{Domain wall operators}
Let the $1D$ Ising model be defined by
\begin{eqnarray}
 S = \sum_i \left[ K s_{i} s_{i+1} + \frac{H}{2} (s_i + s_{i+1})
\right ]\label{SIsing}
\end{eqnarray}
where the temperature factor is absorbed in the symbols $K$ and
$H$. At low temperatures ($K \rightarrow \infty$) the free energy
for the system with $L$ spins becomes simply
\begin{eqnarray}
 F=-\ln Z = -(K+ |H|) L \label{FIsing}
\end{eqnarray}
indicating, like Eq. \eqref{FlargeN}, that a first order
transition occurs at $H=0$. The excitations of lowest energy are
described by the {\em domain walls} separating an array of
up-spins from an array of down-spins. Consider first $H > 0$ such
that the ground state has all the spins upward. Let furthermore
$a_{+-} (j)$ denote the operator that creates a domain wall at
lattice site $j$, i.e. $s_i =+1$ for $i \leq j$ and $s_i = -1$ for
$i>j$. Similarly $a_{-+} (j)$ denotes the domain wall operator
with the $+$ and $-$ spins interchanged. In short hand notation
one can write
\begin{eqnarray}
 \langle a_{+-} (i) a_{-+} (j) \rangle &=& e^{-4K} ~ \vartheta (j-i)
 e^{-2H(j-i)}
 ~~~~~~~~~(H > 0). \label{Hplus}
\end{eqnarray}
In the same way one finds
\begin{eqnarray}
 \langle a_{+-} (i) a_{-+} (j) \rangle &=& e^{-4K} ~ \vartheta (i-j)
 e^{+2H(j-1)}
 ~~~~~~~~~(H < 0). \label{Hminus}
\end{eqnarray}
By using an appropriate definition of the step function
$\vartheta$ on the lattice then these expressions precisely
correspond to those of the Coulomb gas, Eqs \eqref{thetaminus} and
\eqref{thetaplus}, obtained from the $m=0$ and $m=-1$ sector
respectively.
\begin{table}[tbp]
\begin{center}
\caption{Mapping of $1D$ Coulomb gas onto the $1D$ Ising model}
\begin{tabular}{lcc}\hline
Quantity       & Coulomb gas                          & Ising        \\
\hline
Symmetry breaking &                                   &              \\
field          & $\omega >0$ &$H>0$    \\
Fugacity       & $\frac{\sigma}{\beta}$               & $e^{-2K}$\\
Operator       &$\frac{\sigma}{\beta} e^{ i\beta A_0}$& $a_{+-}$ \\
Operator       &$\frac{\sigma}{\beta} e^{-i\beta A_0}$& $a_{-+}$
\\
Partition function & $\left \langle e^{\frac{2\sigma}{\beta} \int
dx \cos \beta A_0 (x)} \right \rangle_{\theta \rightarrow \pi^-}$
& $\left \langle e^{\sum_i \left( a_{+-} (i) + a_{-+} (i)
\right)} \right\rangle_{H \rightarrow 0^+}$ \\
& & \\ \hline
\end{tabular}
\label{TZM1}
\end{center}
\end{table}
Next, one can easily check that all the terms of the low
temperature expansion of the Ising model are in one-to-one
correspondence with those of the series expansion of the Coulomb
gas in powers of $\sigma/\beta$. Considering $H \gtrsim 0$ from
now onward then, under the appropriate identification of
parameters, one can express the four point correlation function in
terms of Eq. \eqref{g4}
\begin{equation}
\langle a_{+-} (i_1) a_{-+} (j_1)  a_{+-} (i_2) a_{-+} (j_2)
\rangle = e^{-8K} g_4
 (i_1 , j_1 , i_2 , j_2 ). \label{fourIsing}
\end{equation}
In the same way one can express the $2n$ point function in terms
of $g_{2n}$, Eq.~\eqref{g2nGen},
\begin{equation}
 \langle a_{+-} (i_1) a_{-+} (j_1) \dots a_{+-} (i_n) a_{-+} (j_n) \rangle =
 e^{-4nK} g_{2n}
 (i_1 , j_1 , \dots , i_n , j_n ) .\label{multiIsing}
\end{equation}
On the basis of the low temperature series one readily concludes
that the partition function of the Ising model can be written as
an operator statement
\begin{equation}
 Z_\textrm{Ising} = \left \langle \exp \sum_i \left( a_{+-} (i) + a_{-+} (i)
 \right)  \right \rangle_{H \rightarrow 0^+} \label{ZIsing}
\end{equation}
which is completely analogous to the statement of Eq. \eqref{ZCG}.
We identify the Ising model quantity $e^{-2K}$ with the fugacity
$\sigma/\beta$ of the Coulomb gas and the role played by $H$ is
the same as $\omega$, see Table~\ref{TZM1}.

Notice that in making the comparison between Eqs \eqref{ZCG} and
\eqref{ZIsing} we have only taken the long ranged parts ($g_{2n}$)
of the Coulomb gas correlations ($G_{2n}$) into account. Although
this point may at first instance be discarded, the presence of
short ranged contributions in $G_{2n}$ will nevertheless result in
slightly different expressions for Eqs \eqref{ZCG} and
\eqref{ZIsing}. These differences will be investigated in a
systematic manner in Section \ref{chiral-coulomb}.
\section{\bf $1D$ Chiral fermions \label{chiralF}}
The simplest formalism that most effectively deals with operator
statements like Eqs \eqref{ZCG} and \eqref{ZIsing} is provided by
none other than the theory of $1D$ chiral fermions. This theory
has previously emerged as the theory of {\em massless} chiral edge
excitations in quantum Hall systems and is otherwise known from
studies on quantum spin chains. The chiral fermion action is given
in terms of fermion fields as follows
\begin{equation}
 S = \int dx\,\bar{\Psi}(x)(i\partial_x + i H\tau_z)
 \Psi(x)\label{SCG0}
\end{equation}
where $\bar\Psi=\{\bar{\Psi}_+,\bar{\Psi}_-\}$,
$\Psi=\{\Psi_+,\Psi_-\}^T$ and $\tau_a$ with $a=x,y,z$ denote the
Pauli matrices. The free energy $F= -\ln Z =-\ln \int
\mathcal{D}[\bar\Psi,\Psi] \exp S$ with varying $H$ is readily
obtained as
\begin{equation}
 F = - \int dx \Tr\ln (i\partial_x + i H\tau_z)=-|H|L.
\end{equation}
This result indicates that chiral fermion theory, like the Coulomb
gas (Eq.~\eqref{FlargeN}) and the Ising model
(Eq.~\eqref{FIsing}), describes a {\em first order} transition as
$H$ passes through zero. Next we consider the expectations ($H
> 0$)
\begin{equation}
 \langle \bar{\Psi}_\pm(x) \Psi_\pm(y)\rangle = \int \frac{d k}{2\pi}
 \frac{e^{ik(x-y)}}{k\pm iH} = \pm i \vartheta(\pm(y-x)) e^{- H |y-x|}.
\end{equation}
To make contact with the results of the Coulomb gas as well as the
Ising model we next introduce the following quantities
\begin{eqnarray}
 a(x) &=& i \bar{\Psi}(x) \frac{\tau_x+i\tau_y}{2}{\Psi}(x) \\
 \bar{a}(x) &=& i \bar{\Psi}(x)\frac{\tau_x-i\tau_y}{2}{\Psi}(x) \label{CFoperators}.
\end{eqnarray}
Notice that under the transformation $H \rightarrow -H$ the
operators $a$ and $\bar{a}$ are interchanged. Assuming $H>0$ from
now onward then the expression for the multi point correlation
function becomes
\begin{equation}
 \langle \bar{a} (x_1) a(y_1)\dots \bar{a} (x_n) a(y_n) \rangle ~=~
 g_{2n}(x_1,y_1,\dots,x_n,y_n)
\end{equation}
which coincides \emph{exactly} with the results found for the
Coulomb gas, Eq.~\eqref{g2nGen}. The operator algebra of the
chiral fermion theory involves one more operator denoted by $Q$
\begin{equation}
 Q = - \frac{i}{\sqrt{2}}\bar{\Psi}\tau_z\Psi.
\end{equation}
Some examples of non-vanishing correlation functions containing
the $Q$ are given by
\begin{eqnarray}
 \langle Q(x) \rangle &=&  \frac{1}{\sqrt{2}} \\
\langle Q(x) \bar{a}(y) a(z)\rangle &=&
\frac{1}{\sqrt{2}}\vartheta(z-y) e^{-2H(z-y)}\notag\\
&+&\frac{1}{\sqrt{2}} \vartheta(x-z)\vartheta(y-z)\vartheta(y-x)
e^{-2H(y-x)}.
\end{eqnarray}
\subsection{Tentative solution of the Coulomb gas problem \label{tentative}}
To see how the chiral fermion action elucidates the complete
singularity structure of the theory we proceed and map the
operator statement of Eq. \eqref{ZIsing} directly onto the theory
of chiral fermions. The relation between the Ising model and
chiral fermion theory is given by
\begin{equation}
\sum_i ( a_{+-}(i) + a_{-+}(i) )
 \leftrightarrow e^{-2K} \int dx ( \bar{a}(x) + a(x) )
 = -i e^{-2K}
 \int dx \bar{\Psi}(x)\tau_x\Psi(x).
\end{equation}
The effective action for the Ising model at finite but low
temperatures can therefore be written as follows
\begin{equation}
 S_\textrm{Ising} = \int dx \bar{\Psi}(x) (i\partial_x + i H\tau_z+ie^{-2K} \tau_x )
  \Psi(x).\label{SIsing-eff}
\end{equation}
This theory is solved in a trivial manner. Introducing an
orthogonal rotation $U=\exp(i\phi\tau_y)$ on the $\bar{\Psi} ,
\Psi$ fields
\begin{equation}
\chi = e^{i\phi \tau_y}\Psi,\qquad \phi =
 \frac{1}{2} \arcsin \frac{e^{-2K}}{\sqrt{H^2 +
 e^{-4K}}}\label{varphi}
\end{equation}
then the action is diagonal
\begin{equation}
 S_\textrm{Ising} = \int dx \bar{\chi}(x) (i\partial_x + i\tilde{H}\tau_z )
 \chi(x).\label{SIsing-eff1}
\end{equation}
Here, the $\tilde{H}$ is defined by
\begin{equation}
 \tilde{H} = \sqrt{H^2 + e^{-4K}}
 \label{masses}
\end{equation}
indicating, as is well known, that the Ising system at finite
temperatures displays a finite mass gap of size $e^{-2K}$. As a
final remark, it should be mentioned that the chiral fermion
action of Eq. \eqref{SIsing-eff} does not yet provide the complete
answer to the Coulomb gas problem. The main reason is that the
correlation functions $G_{2n}$ of the Coulomb gas have short
distance contributions that are generally different from the
asymptotic scaling form that we have denoted by $g_{2n}$. We will
embark on the subtle differences between the Ising model and the
Coulomb gas in Section \ref{chiral-coulomb}.
\subsection{Comparison with Ising model at finite temperatures}
\subsubsection{Magnetization \label{magnetization}}
Let us next compare the predictions based on the chiral fermion
action of Eq. \eqref{SIsing-eff1} with the exact solutions of the
Ising model. For example, the magnetization per spin is obtained
as follows
\begin{equation}
 \mathcal{M} = -\frac{1}{L}\frac{\partial F}{\partial H}
 = \frac{H}{\sqrt{H^2 + e^{-4K}} }
\end{equation}
which for small values of $H$ precisely corresponds to the exact
result~\cite{Baxter}
\begin{equation}
 \mathcal{M} = \frac{\sinh H}{\sqrt{\sinh^2 H + e^{-4K}}
 }.\label{MagnetIsing}
\end{equation}
Notice that the spontaneous magnetization vanishes everywhere
except at zero temperature ($K \rightarrow \infty$) as it should
be.
\subsubsection{Domain wall operators}
Of interest next are the Ising model two point correlations at
finite temperatures. For this purpose we make use of the
orthogonal rotation $U$ discussed in the previous Section and
express the local operators $\bar{a}$, $a$ and $Q$ in terms of the
quantities
\begin{eqnarray}
 \tilde{a}(x) &=& i \bar{\chi}(x) \frac{\tau_x+i\tau_y}{2}{\chi}(x) \\
 \tilde{\bar{a}}(x) &=& i \bar{\chi}(x)\frac{\tau_x-i\tau_y}{2}{\chi}(x)\\
  \tilde{Q} &=& - \frac{i}{\sqrt{2}}\bar{\chi}\tau_z \chi . \label{CFoperators2}
\end{eqnarray}
for which the correlation functions are simple. The relation
between the two different sets of operators can be written as
follows
\begin{eqnarray}
 \left(
\begin{matrix}
 \bar{a} \\ a \\ Q
\end{matrix}
 \right) =
 \left(
\begin{matrix}
 \cos^2 \phi & -\sin^2 \phi & -\sqrt{2} \sin \phi \cos \phi \\
 -\sin^2 \phi & \cos^2 \phi & -\sqrt{2} \sin \phi \cos \phi \\
 \sqrt{2} \sin \phi \cos \phi & \sqrt{2} \sin \phi \cos \phi &
 \cos^2 \phi -\sin^2 \phi
\end{matrix}
 \right)
 \left(
\begin{matrix}
 \tilde{\bar{a}} \\ \tilde{a} \\ \tilde{Q}
\end{matrix}
 \right) .\label{newoperators}
\end{eqnarray}
On the basis of Eq. \eqref{newoperators} we obtain the following
expression for the pair correlation
\begin{equation}
g_2(x,y)= \langle \bar{a} (x) a(y) \rangle
 =  m_0 +  m_+  \vartheta(y-x) e^{-\tilde{H} |y-x|} + m_-
 \vartheta (x-y) e^{-\tilde{H} |y-x|} \label{adaga}
\end{equation}
where
\begin{eqnarray}
 m_0 &=& \frac{1}{4} \left( 1 - \mathcal{M}^2 \right) \label{m0}\\
 m_\pm &=& \frac{1}{4} \left(1 \pm \mathcal{M}  \right)^2 \label{m+}\\
 \tilde{H} &=& ~~~~~ \frac{H}{\mathcal{M}} .\label{mmm}
\end{eqnarray}
It is a matter of simple algebra to show that the results of this
Section, which include the orthogonal rotation $U$, are all in
one-to-one correspondence with those obtained from the standard
transfer matrix approach to the Ising model (see Appendix).
\subsection{Ising model mass gap}
Before proceeding with the details of the mapping, it is helpful
to first digress on the meaning of the Ising model mass gap
$\tilde{H} = e^{-2K}$ at $H=0$ (see Eq. \ref{masses}). Even though
mass generation may generally be regarded as one of most
interesting aspects of the $1D$ Ising model, from the Coulomb gas
or $\theta$ vacuum point of view the meaning of this phenomenon is
very different, however, and this so because of the difference in
dimensionality. Notice that in the language of the Coulomb gas
(Table 1) the mass gap at $\theta=\pi$ is solely induced by the
fugacity $\sigma/\beta$ in the problem and both quantities vanish
in the limit where the linear dimension $\beta$ goes to infinity
which is the limit of physical interest. Therefore, the large $N$
expansion at $\theta=\pi$ is fundamentally {\em gapless} and the
results sofar indicate that the sought after critical theory is
the same as that of the $1D$ Ising model at low temperatures or,
equivalently, the theory of chiral fermions.

To further elucidate meaning of the correlation functions of Eqs
\eqref{adaga} - \eqref{mmm} for the Coulomb gas at $\theta=\pi$ we
next consider the {\em critical} correlations for which the
exponential factor $e^{-\tilde{H}|x-y|}$ is close to unity. This
means that we take the $H$ field to be close to zero (or $\theta
\approx \pi$) and, at the same time, the distance $|x-y|$ to be
much smaller than the Ising model correlation length $e^{2K}$ (or
$\beta/\sigma$). After simple algebra it follows directly that Eq.
\eqref{adaga} can be written in the general form
\begin{equation}
 \langle\bar{a} (x) a(y)\rangle =  e^{- {2\tilde{H}} |y-x|/ \mathcal{M}}
 \left\{ \tilde{\mathcal{P}}_\pi
 + \tilde{\mathcal{P}}_0 ~ \vartheta(y-x)
 + \tilde{\mathcal{P}}_{2\pi} ~ \vartheta (x-y) \right\} \label{finiteT}
\end{equation}
where
\begin{eqnarray}
\tilde{\mathcal{P}}_{0} &=&1- \tilde{\sigma}_{xy} -
 \tilde{\sigma}_{xx} \\
 \tilde{\mathcal{P}}_{\pi} &=& 2  \tilde{\sigma}_{xx} \\
  \tilde{\mathcal{P}}_{2\pi} &=&  \tilde{\sigma}_{xy} -
 \tilde{\sigma}_{xx}
\end{eqnarray}
and
\begin{eqnarray}
 \tilde{\sigma}_{xy} &=& (m_- + m_0)
 = \frac{1}{2} (1-\mathcal{M})  \label{tildexy}\\
 \tilde{\sigma}_{xx} &=& m_0 \left( e^{{2H} |y-x| /\mathcal{M}} -1
 \right) \approx
 |x-y| \frac{e^{-4K}}{\sqrt{ H^2 + e^{-4K}}} \ll 1. \label{tildexx}
\end{eqnarray}
These expressions are completely analogous to those of the free
energy of the Coulomb gas with finite values of $\beta$ and $L$
and varying boundary conditions, see Eq. \eqref{free}. The
quantities $\tilde{\sigma}_{xx}$ and $\tilde{\sigma}_{xy}$ in Eqs
\eqref{tildexy} and \eqref{tildexx} have the same meaning as the
conductance parameters $\langle\sigma_{xx}^\prime\rangle$ and
$\langle\sigma_{xy}^\prime\rangle$ in Eq. \eqref{free}. Their
detailed dependence on $H$ or $\theta$ is very different, however,
and these differences reflect the distinctly different ways in
which the infrared of the system is being regulated in each case.

In summary, based on Eqs \eqref{finiteT} - \eqref{tildexx} we can
say that the ``conductance" parameters
$\langle\sigma_{xx}^\prime\rangle$ and
$\langle\sigma_{xy}^\prime\rangle$ quite generally reveal
themselves as the most important physical observables of the large
$N$ expansion. Notice that these quantities naturally emerge from
the Coulomb gas provided the {\em infrared} of the system is
properly defined, e.g. either by taking the linear dimension $L$
to be finite as in Eq. \eqref{free}, or by working with finite
values of the fugacity $\sigma/\beta$ as in Eqs \eqref{finiteT} -
\eqref{tildexx}. This means that
$\langle\sigma_{xx}^\prime\rangle$ and
$\langle\sigma_{xy}^\prime\rangle$ are the fundamental objects of
the theory in which the super universal strong coupling features
of the $\theta$ vacuum can generally be expressed. Finally, the
results of Eqs \eqref{finiteT} - \eqref{tildexx} explain, at the
same time, why Coleman's picture of the transition at
$\theta=\pi$, which is based on the zero temperature expressions
of Eqs~\eqref{thetaminus} and \eqref{thetaplus} alone, is in many
ways too simple. This picture all by itself does not facilitate a
correct analysis of the topological features of an ``edge", in
particular the appearance of an edge spin or edge currents, nor
does it recognize the existence of physical quantities like
$\langle\sigma_{xx}^\prime\rangle$ and
$\langle\sigma_{xy}^\prime\rangle$ that generally provide the most
important information on the low energy dynamics of the $\theta$
vacuum. Given this lack of insight in the infrared properties that
one generally can associate with the topological issue of an
instanton vacuum, it may no longer be any surprise to know that
the concept of super universality has historically been overlooked
completely~{\cite{Affleck85,Affleck88}}.
\section{Mapping of $1D$ Coulomb gas onto $1D$ chiral fermions
\label{chiral-coulomb}}
\subsection{Finite renormalizations}
In this Section we complete the mapping of the Coulomb gas onto
chiral fermions and address the {\em short distance} parts of the
correlations $G_{2n}$ which so far have been ignored. These short
distance parts can in general be easily separated from the long
distance contributions that we have denoted by $g_{2n}$. The basic
idea therefore is to proceed by {\em eliminating} the short
distance correlations from the Coulomb gas problem of Eq.
\eqref{ZCG} while retaining all the long distance parts $g_{2n}$.
This procedure can in principle be carried out order by order in a
series expansion of Eq. \eqref{ZCG} in powers of the fugacity
$\sigma/\beta$. The aim of this procedure is to eventually express
the Coulomb gas problem in terms of the pure scaling operators
$\bar{a}$ and $a$ of the chiral fermion theory, rather than the
original charge operators $e^{\pm i \beta A_0}$. Since this
elimination procedure generally involves lengthy but elementary
computations we shall proceed by first quoting the final results.
Then, instead of embarking on the details of the computations we
shall, in Section \ref{hamil}, present a more elegant and
effective computational scheme based on the hamiltonian approach.

Assuming as in Eq. \eqref{ZCG} that $\theta$ approaches $\pi$ from
below then the mass term $2\frac{\sigma}{\beta} \cos \beta A_0$ of
the Coulomb gas problem can be expressed in terms of the pure
scaling operators $\bar{a}$ and $a$ according to
\begin{equation}
 \left \langle e^{\frac{2\sigma}{\beta} \int dx \cos\beta A_0} \right
 \rangle_{\theta \rightarrow \pi^-} \equiv
\left \langle e^{\frac{2\sigma}{\beta} \int dx \cos\beta A_0}
\right
 \rangle_\omega
 =  \left \langle e^{\frac{\sigma}{\beta}
 \int dx ( Z \bar{a} + Z a + Z_0)}\right
 \rangle_{Z_\omega \omega}. \label{expanding}
\end{equation}
Here, the expectations are defined for the theory with fugacity
zero and the limit $L \rightarrow \infty$ is understood. Eq.
\eqref{expanding} tells us that the aforementioned elimination
process generally involves three distinct renormalization
coefficients, i.e., one for the $\omega$ field $\omega \rightarrow
Z_\omega \omega$, a second one for the $\sigma$ variable $\sigma
\rightarrow Z \sigma$ and a third one, $Z_0$, which is a constant.
The quantities $Z_\omega$, $Z$ and $Z_0$ are regular functions for
small values  of $\omega$ and can be expressed in terms of a
regular series expansion in powers of $\sigma/\beta$. To
illustrate the procedure we expand the left hand side of Eq.
\eqref{expanding} in powers of $\sigma/\beta$. To lowest
non-trivial order we obtain the following terms
\begin{equation}
 \left( \frac{\sigma}{\beta} \right)^2 \int dx_1 \int dx_2\,
 G_2(x_1,x_2)
 = \left( \frac{\sigma}{\beta} \right)^2 \int dx_1 \left[
 \int dx_2\, g_2 (x_1,x_2) + \frac{4 \kappa}{1+\theta/\pi} \right] .
 \label{contri2}
\end{equation}
Here, the short ranged term in $G_2(x_1,x_2)$, Eq.
\eqref{thetaminusfull}, has been integrated out explicitly. After
re-exponentiation of Eq. \eqref{contri2} we immediately obtain the
right hand side of Eq. \eqref{expanding} with $Z=Z_\omega =1$ and
\begin{equation}
 Z_0 = \left( \frac{\sigma\kappa}{\beta} \right)
 \frac{4}{1+\theta/\pi}
\label{Z0}
\end{equation}
By continuing along the same lines but now taking the higher order
terms in $\sigma/\beta$ into account one finds
\begin{eqnarray}
 Z_\omega &=& 1+ \left( \frac{\sigma\kappa}{\beta} \right)^2
 \frac{32}{(1+\theta/\pi)(3-\theta/\pi) }
\label{Zomega} \\
 Z &=& 1 - \left( \frac{\sigma\kappa}{\beta}\right )^2\left [
 \frac{32}{(1+\theta/\pi)(3-\theta/\pi) } +
 \frac{8}{(1+\theta/\pi)^2} +\frac{8}{(3-\theta/\pi)^2}\right ]
 .\label{Z}
\end{eqnarray}
By comparing Eq. \eqref{expanding} with the operator statements of
chiral fermion theory, Section \ref{chiralF}, one can say that the
charged particle operators $e^{\pm i\beta A_0}$ of the Coulomb gas
define an operator algebra that generally involves three critical
operators only, namely the quantities $\bar{a}$ and $a$ that are
associated with the fugacity $\sigma/\beta$ and a distinctly
different operator denoted by $Q$ that is associated with
$\omega$.

We next extend the result of Eq. \eqref{expanding} to include the
expectations of the Coulomb gas operators $e^{\pm i\beta A_0}$ but
now for the theory with finite values of the fugacity. Using the
same notation as before we can write
\begin{equation}
 \left \langle e^{\frac{2\sigma}{\beta} \int dx \cos\beta A_0}
 ~ e^{i\beta A_0 (x)} \right \rangle_\omega
 = \left \langle e^{ \frac{\sigma}{\beta}
 \int dx ( Z \bar{a} + Z a + Z_0) } \left( Z \bar{a} (x) + Z_0 \right)
\right  \rangle_{Z_\omega \omega} \label{expanding1}
\end{equation}
and a similar result for $e^{- i\beta A_0}$. For the pair
correlation one finds
\begin{eqnarray}
&&  \left \langle e^{\frac{2\sigma}{\beta} \int dx \cos\beta A_0}
~ e^{i\beta A_0 (x) - i\beta A_0 (y)}\right
 \rangle_\omega \notag\\
 && \hspace{2cm}= \left \langle e^{ \frac{\sigma}{\beta}
 \int dx ( Z \bar{a} + Z a + Z_0) }
 \left( Z \bar{a} (x) + Z_0 \right) \left( Z a (y) + Z_0 \right)
 \right \rangle_{Z_\omega \omega}.\label{expanding2}
\end{eqnarray}
Eqs \eqref{expanding1} and \eqref{expanding2} involve the same
coefficients $Z$, $Z_\omega$ and $Z_0$ as those obtained before,
i.e., Eqs \eqref{Z0},\eqref{Zomega} and \eqref{Z}. We can
therefore replace the charge operators $e^{\pm i\beta A_0}$ of the
Coulomb gas by the pure scaling operators $\bar{a}$ and $a$
following the general rule (see Table~\ref{TZM})
\begin{eqnarray}
 e^{i\beta A_0 (x)} &\rightarrow& Z \bar{a} (x) + Z_0 \nonumber \\
 e^{-i\beta A_0 (x)} &\rightarrow& Z {a} (x) + Z_0  .
 \label{expanding3}
\end{eqnarray}
Evaluation of Eq.~\eqref{expanding2} yields the following result
for pair correlation function
\begin{equation}
 \left \langle e^{i\beta A_0 (x) - i\beta A_0 (y)} \right
 \rangle_{CG}= M_0 + M_+ \vartheta(y-x) e^{-2\tilde\omega|y-x|}+ M_-
\vartheta(x-y) e^{-2\tilde\omega|y-x|}\label{G2FULL}
\end{equation}
where
\begin{eqnarray}
M_0 &=& \left (\frac{1}{2} Z \sqrt{1-\tilde{\mathcal{M}^2}} + Z_0\right )^2\\
M_\pm &=&  \frac{Z^2}{4}\left (1\pm \tilde{\mathcal{M}}\right )^2 \\
\tilde{\mathcal{M}} &=& \frac{Z_\omega
 \omega}{\sqrt{Z_\omega^2 \omega^2
 + Z^2 \sigma^2 / \beta^2}}.
\end{eqnarray}

\begin{table}[tbp]
\begin{center}
\caption{Critical operators in Coulomb gas representation, chiral
fermion theory and the $1D$ Ising model}
\begin{tabular}{lccc}\hline
Operator & Coulomb gas
& Chiral fermions & $1D$ Ising \\
\hline a      & $e^{ i\beta A_0} =Z\bar{a} + Z_0$ & $i
\bar{\Psi}\frac{\tau_x+i\tau_y}{2} \Psi$ & $a_{+-}
= \frac{\tau_x+i\tau_y}{2}$ \\
       &                   &                         &          \\
$\bar a$ & $e^{-i\beta A_0} = Z a + Z_0$ & $i \bar{\Psi}
\frac{\tau_x-i\tau_y}{2}\Psi$ & $a_{-+}
= \frac{\tau_x-i\tau_y}{2}$ \\
       &                   &                         &          \\
$Q$      & $Z_\omega Q$              & $-\frac{i}{\sqrt{2}}
\bar{\Psi}\tau_z \Psi $ &
$s = \frac{1}{\sqrt{2}}\tau_z$        \\
\hline
\end{tabular}
\label{TZM}
\end{center}
\end{table}

%
\subsection{Hamiltonian approach \label{hamil}}
In this Section we show how ordinary quantum mechanics can be used
very effectively to compute the various different numerical
aspects of Coulomb gas problem, in particular the coefficients
$Z$, $Z_\omega$ and $Z_0$. For this purpose we consider the
hamiltonian of the $1D$ Coulomb gas action, Eq. \eqref{Zmlmr},
which for infinite systems ($L \to \infty$) can be written as
\begin{equation}
\mathcal{H} = \frac{1}{4\kappa} \left (-i \frac{\partial}{\partial
(\beta A_0)} -\frac{\theta}{2\pi}\right )^2
-2\frac{\sigma}{\beta}\cos(\beta A_0).\label{H1}
\end{equation}
This hamiltonian acts on wave functions with periodic boundary
conditions $\psi(\beta A_0+2\pi)=\psi(\beta A_0)$. In the limit of
zero fugacity $\sigma/\beta=0$ the eigenvalues and eigenfunctions
of $\mathcal{H}$ are easily found
\begin{equation}
E^{(0)}_m = \frac{1}{4\kappa}\left(m+\frac{\theta}{2\pi}\right
)^2,\qquad \psi^{(0)}_m = \frac{1}{\sqrt{2\pi}} e^{-i m \beta
A_0}, \qquad m \in \mathbb{Z}.
\end{equation}
Notice that the energy levels of the $m=0$ and $m=-1$ sectors
cross one another at $\theta=\pi$. The $(2\sigma/\beta)\cos\beta
A_0$ term in Eq. \eqref{H1} produces a band splitting which can be
dealt with using standard perturbation theory.

We consider the two-point function $G_2(x,y)$
\begin{equation}
G_2(x,y) =
  \begin{cases}
    \sum\limits_{J=0}^\infty |\langle 0|e^{-i\beta A_0}
    |J\rangle |^2 e^{-(E_J-E_0)|y-x|} & \qquad
    x\leq y, \\
    \sum\limits_{J=0}^\infty |\langle J |e^{-i\beta A_0}
    |0\rangle |^2 e^{-(E_J-E_0)|y-x|} & \qquad x > y.
  \end{cases} \label{G2Ham}
\end{equation}
Here $E_J$ and $|J\rangle $ denote the exact eigenvalues and
eigenstates respectively. The long range part of $G_2(x,y)$ is
determined by the terms in Eq. \eqref{G2Ham} with $J=0,1$,
\begin{eqnarray}
 g_2(x,y) = |\langle 0|e^{-i\beta A_0} |0\rangle |^2 &+& |\langle
 0|e^{-i\beta A_0} |1\rangle |^2 \vartheta(y-x) e^{-(E_1-E_0)|y-x|}
\notag \\
&+& |\langle 1|e^{-i\beta A_0} |0\rangle |^2 \vartheta(x-y)
 e^{-(E_1-E_0)|y-x|} .\label{g2Ham}
\end{eqnarray}
To study the limit $\theta \to \pi^{-}$ we proceed by first
projecting the hamiltonian onto the subspace of eigenfunctions
$\psi^{(0)}_0$ and $\psi^{(0)}_{-1}$. The following estimates for
two lowest energies are obtained
\begin{equation}
E_{0,1}^{(1)} = \frac{1}{4\kappa}\left (\frac{\theta}{2\pi}\right
)^2+\omega \mp \sqrt{\omega^2 + (\sigma/\beta)^2}.
\end{equation}
The corresponding eigenfunctions are
\begin{equation}
   \begin{pmatrix}
    \psi_0^{(1)}  \\
    \psi_{-1}^{(1)}
  \end{pmatrix} = e^{i\phi \tau_y}
  \begin{pmatrix}
    \psi^{(0)}_0  \\
    \psi^{(0)}_{-1}
  \end{pmatrix} .
\end{equation}
Here, $\omega$  is defined by Eq. \eqref{omega} and the angle
$\phi$ by Eq.~\eqref{varphi}. Using these results we obtain the
same expression for $g_2$ as in Eq. \eqref{adaga}, i.e.,
\begin{equation}
g_2(x,y) = m_0 + m_+ \vartheta(y-x) e^{-2\tilde\omega|y-x|}+ m_-
\vartheta(x-y) e^{-2\tilde\omega|y-x|}.\label{adagaHam}
\end{equation}
with $\tilde\omega = \sqrt{\omega^2 + (\sigma/\beta)^2}$. The
quantities $m_0$ and $m_{\pm}$, under the appropriate substitution
of parameters (see Table~\ref{TZM1}), are given by
Eqs~\eqref{m0}-\eqref{m+}.

From this point onward we use perturbation theory in the fugacity
$\sigma/\beta$, taking into account the presence of the other
levels. The results when compared to the general expression of Eq.
\eqref{G2FULL} can be used to extract the coefficients $Z$, $Z_0$
and $Z_\omega$. For example, for the energy gap between the first
excited state and the ground state we obtain to next leading order
in $\sigma/\beta$
\begin{equation}
E_{1}^{(1)} - E_0^{(1)} = 2 \sqrt{\omega^2 Z_\omega^2 +
(\sigma/\beta)^2}
\end{equation}
where $Z_\omega$ is given by Eq.~\eqref{Zomega}. Similarly, the
eigenfunctions to next leading order are given by
\begin{eqnarray}
\psi_0^{(2)} &=& \left [1- \left(\frac{4\kappa\sigma}{\beta}\right
)^2 \frac{1}{2(1+\theta/\pi)^2} \right ]\psi_0^{(1)} +
\frac{4\kappa\sigma}{\beta}\frac{\sin\phi}{2(2-\theta/\pi)}\psi_{-2}^{(0)}\\
&&+\frac{4\kappa\sigma}{\beta}\frac{\cos\phi}{1+\theta/\pi}\psi_{1}^{(0)}
+ \left(\frac{4\kappa\sigma}{\beta}\right
)^2\frac{1}{2(1+\theta/\pi)(2+\theta/\pi)}\psi_{2}^{(0)}\label{10}
\\
\psi_{-1}^{(2)} &=& \left
[1-\left(\frac{4\kappa\sigma}{\beta}\right
)^2\frac{1}{2(3-\theta/\pi)^2} \right ]\psi_{-1}^{(1)} +
\frac{4\kappa\sigma}{\beta}\frac{\cos\phi}{3-\theta/\pi}\psi_{-2}^{(0)}
\\
&&-\frac{4\kappa\sigma}{\beta}\frac{\sin\phi}{2\theta/\pi}\psi_{1}^{(0)}
+ \left(\frac{4\kappa\sigma}{\beta}\right
)^2\frac{1}{2(3-\theta/\pi)(4-\theta/\pi)}\psi_{3}^{(0)}\label{11}
\end{eqnarray}
Using these expressions we find for the matrix elements in Eq.
\eqref{g2Ham}
\begin{equation}
 |\langle 0|e^{-i\beta A_0} |0\rangle |^2 = \left (\frac{Z}{2}\sin
 2\phi +Z_0\right )^2,\qquad |\langle 0|e^{-i\beta A_0}
 |1\rangle |^2 = Z^2 \cos^4\phi
\end{equation}
where the quantities $Z_0$ and $Z$ are given by Eqs.~\eqref{Z0}
and \eqref{Z} respectively.
\section{Finite size systems \label{finite}}
\subsection{Introduction}
We have now completed one of the main objectives of this paper
which is to lay the bridge between the large $N$ expansion or
Coulomb gas near $\theta = \pi$ and exactly solvable models in one
dimension. We next wish to extend this mapping to include the
Coulomb gas with varying linear dimensions $\beta$ and $L$ which
defines the conductances $\langle\sigma_{xx}^\prime \rangle$ and
$\langle\sigma_{xy}^\prime \rangle$. For this purpose we will
study spin chains and chiral fermion theory with finite values of
$L$. By expressing the conductances in terms of both Ising model
and chiral fermion correlations we are able to extend the
previously obtained scaling results of Eqs~\eqref{kubo1} and
\eqref{kubo2} to include infinite orders in the fugacity
$\sigma/\beta$. The final expressions that we obtain have the
appropriate behavior in the thermodynamic limit $L,\beta \to
\infty$ and serve as a starting point in Section \ref{RGLN} where
we discuss the renormalization behavior of the Coulomb gas.
\subsection{$1D$ Ising model (I) \label{ising1}}
To start we consider the partition function of an Ising spin chain
of length $L$. In terms of the transfer matrix
\begin{equation}
 T = e^K \left (\cosh H + \tau_z \sinh H + \tau_x e^{-2K}\right
 )\label{TransfMat}
\end{equation}
we write
\begin{equation}
 Z = \Tr T^L B .\label{finiteIsing}
\end{equation}
Here, the $2\times 2$ matrix B defines the boundary conditions on
the spin chain. For example, $B=\bf{1}$ corresponds to {\em
periodic} boundary conditions, $B=\tau_x$ describes {\em twisted}
boundary conditions and $B = \cosh H + \tau_z \sinh H + \tau_x$
corresponds to {\em free} (or {\em no}) boundary conditions.

The idea next is to find the explicit form for $B$ such that Eq.
\eqref{finiteIsing} can be identified with the partition function
of the Coulomb gas, Eqs \eqref{free}. After some investigation it
is not difficult to see that the correct expression for the matrix
$B$ is given by
\begin{equation}
 B(q,\beta a_0) =
\begin{pmatrix}
 1 & e^{i \pi q + i \beta a_0} \\ e^{i \pi q - i \beta a_0} & e^{2i \pi q }
\end{pmatrix} \label{matrix}
\end{equation}
and the partition function of the Coulomb gas, Eqs \eqref{free},
can be obtained as
\begin{eqnarray}
 Z[q, \beta a_0] = \frac{\Tr T^L B(q, \beta a_0) }{ \Tr T^L B(0,0)} .
 \label{free1aa}
\end{eqnarray}
To show this we consider the low temperature limit $K \rightarrow
\infty$. To lowest non-trivial order in an expansion in powers of
$e^{-2K}$ we can write Eq. \eqref{free1aa} in the general form
\begin{eqnarray}
 Z[q, \beta a_0]
 &=&  \left (1-{\langle \sigma_{xy}^\prime \rangle} -
 \langle \sigma_{xx}^\prime \rangle \right ) e^{0 i q}
 +   2  \langle \sigma_{xx}^\prime \rangle
 e^{ \pi i q} \cos \beta a_0 \notag \\ &&+
\left ({\langle \sigma_{xy}^\prime \rangle}-
 \langle \sigma_{xx}^\prime \rangle \right )
 e^{2 \pi i q}.
 \label{free1a}
\end{eqnarray}
Here, $\langle \sigma_{xx}^\prime \rangle$ and $\langle
\sigma_{xy}^\prime \rangle$ are given by the same expressions as
in Eqs~\eqref{kubo1} and \eqref{kubo2}, i.e.,
\begin{eqnarray}
 \langle \sigma_{xx}^\prime \rangle &=&  \frac{\eta}{e^{X} + e^{-X} +2\eta }
 \label{diss0}\\
 \langle \sigma_{xy}^\prime \rangle &=&
 \frac{1}{2} \left[  1 -
 \frac{e^{X} - e^{-X}}{e^{X} + e^{-X} +2\eta  }
 \right] \label{hall0}.
\end{eqnarray}
The quantities $X$ and $\eta$ now defined as
\begin{eqnarray}
 X &=& L H  \label{XetaIsing1} \\
 \eta &=& L e^{-2K} \frac{\sinh LH}{L \sinh H}
 \rightarrow L e^{-2K} \frac{\sinh X}{X} \label{XetaIsing2}
\end{eqnarray}
where in the last step we have taken the limit of small $H$. By
comparing Eqs~\eqref{XetaIsing1} and \eqref{XetaIsing2} with Eqs
\eqref{XetaCG1} and \eqref{XetaCG2} we conclude that under the
appropriate substitution of variables (see Table ~\ref{TZM1}) the
mapping of the Coulomb gas problem onto the $1D$ Ising model is
retained also for finite size $L$.
\subsection{1D Ising model (II) \label{ising2}}
We now can proceed and perform the summation of the series in
powers of $e^{-2K}$ to infinite order by employing the orthogonal
rotation $U$ introduced in Section \ref{tentative}. More
specifically, in the limit of small $e^{-2K}$ and $H$ we can write
the transfer matrix $T$ as follows
\begin{equation}
 T=e^{2K} U^{-1} e^{\tilde{H}\tau_z} U.
\end{equation}
The quantity $Z[q,\beta a_0]$ can therefore be expressed as
\begin{eqnarray}
 Z[q, \beta a_0] = \frac{\Tr e^{L \tilde{H}\tau_z} U B(q, \beta a_0)U^{-1} }
 {\Tr e^{L \tilde{H}\tau_z} U B(0,0)U^{-1} } .
\end{eqnarray}
The result is of the same general form as Eq. \eqref{free1a} but
with the conductances $\langle \sigma_{xx}^\prime \rangle $ and
$\langle \sigma_{xy}^\prime \rangle $ now given as
\begin{eqnarray}
 \langle \sigma_{xx}^\prime \rangle &=&  \frac{\eta}{e^{X} + e^{-X} +2\eta }
 \label{diss}\\
 \langle \sigma_{xy}^\prime \rangle &=&
 \frac{1}{2} \left[  1 -
 \frac{ \sqrt{  (e^{X} - e^{-X})^2 - 4 \eta^2 }}{e^{X} + e^{-X} +2\eta  }
 \right] = \frac{1}{2} \left[  1 -
 \mathcal{M} \frac{ {e^{X} - e^{-X}}}{e^{X} + e^{-X} +2\eta  }
 \right]\label{hall}
\end{eqnarray}
where
\begin{eqnarray}
 X &=& L \tilde{H} = \sqrt{H^2 + e^{-4K}} \label{Xweak}\\
 \eta &=& \sqrt{1-\mathcal{M}^2}\sinh X  =
 L e^{-2K} \frac{\sinh X}{X} .
 \label{etaweak}
\end{eqnarray}
Notice that these expressions cannot be obtained from the lowest
order results of Eqs \eqref{diss0} - \eqref{XetaIsing2} by
considering the series expansion to any finite order in powers of
$e^{-2K}$. In the language of the Coulomb gas we can say that the
higher order contributions in the fugacity $\sigma/\beta$ actually
diverge as $\theta$ approaches $\pi$ and the infinite order
results of Eqs \eqref{diss} - \eqref{etaweak} are therefore
qualitatively different from the lowest order ones. For example,
unlike Eqs \eqref{diss0} - \eqref{XetaIsing2} we can now consider
the limit $L \to \infty$ keeping $e^{-2K}$ or $\sigma/\beta$
finite and the result is
\begin{eqnarray}
 \langle \sigma_{xx}^\prime \rangle &=&  \frac{1}{2}
 \frac{\sqrt{1-\mathcal{M}^2}}{1+\sqrt{1-\mathcal{M}^2} }
 \label{diss1a}\\
 \langle \sigma_{xy}^\prime \rangle &=&
  \frac{1}{2} \left[  1 -
 \frac{ { \mathcal{M} }}{1+ \sqrt{1-\mathcal{M}^2} }
 \right] .\label{hall1a}
\end{eqnarray}
These expressions are well behaved at $H=0$ (or $\theta=\pi$) and
very similar to those obtained from the correlation functions, Eqs
\eqref{tildexy} and \eqref{tildexx}. Moreover, by taking $e^{-2K}$
or $\sigma/\beta$ to zero we obtain a sharp transition between the
quantum Hall plateaus, i.e. $\langle \sigma_{xx}^\prime \rangle =
\langle \sigma_{xy}^\prime \rangle =0$ for $H>0$ ($\theta < \pi$)
and $\langle \sigma_{xx}^\prime \rangle = 0 $, $\langle
\sigma_{xy}^\prime \rangle =1$ for $H<0$ ($\theta > \pi$) as it
should be.

From Eqs \eqref{diss1a} and \eqref{hall1a} we furthermore conclude
that the role of the fugacity $\sigma/\beta$ in the Coulomb gas
problem or large $N$ expansion is very similar to that of the
temperature or external frequencies in the theory of localization
and interaction effects. For example, the results of Eqs
\eqref{diss1a} and \eqref{hall1a} are independent of the linear
dimension $L$ of the system and effectively describe the
conductances of a ``finite" sample with linear dimensions $\beta$
and $L = \xi = \beta/\sigma$ respectively. The consequences of Eqs
\eqref{diss} - \eqref{etaweak} for the Coulomb gas problem will be
discussed further in Section \ref{explicit}.
\subsection{$1D$ Ising model (III) \label{ising3}}
Although the one-to-one correspondence between the Coulomb gas and
the $1D$ Ising model is limited to the low temperature regime
$e^{-2K} \rightarrow 0$ of the latter only, it is nevertheless
instructive to express the definition of $Z[q, \beta a_0]$, Eq.
\eqref{free1aa}, quite generally in terms of the Ising model
parameters $K$ and $H$. For this purpose, write $T^L$ in the form
\begin{equation}
 \left( T^L \right)_{\sigma \sigma^\prime} = \exp \left [ -F + K^\prime
 \sigma \sigma^\prime + \frac{H^\prime}{2} (\sigma + \sigma^\prime)
 \right ].
\end{equation}
Here
\begin{eqnarray}
 K^\prime &=& -\ln \frac{\eta}{2} \label{Kprime}\\
 H^\prime &=& ~~\ln \frac{\eta}{2} + \ln \left
 (\frac{1}{2}+\sqrt{1-\mathcal{M}^2}\frac{\eta}{\mathcal{M}}
 \right)\label{Hprime}
\end{eqnarray}
can be taken as the {\em effective} Ising model parameters of a
chain of length $L$. The results can again be written in the
general form of Eq. \eqref{free1a} but with the parameters
$\langle\sigma_{xx}^\prime\rangle$ and
$\langle\sigma_{xy}^\prime\rangle$ now given by
\begin{eqnarray}
 \langle\sigma_{xx}^\prime\rangle &=& \frac{ e^{-2K^\prime}}{e^{H^\prime} + e^{-H^\prime} +
 2e^{-2K^\prime}} \\
 \langle\sigma_{xy}^\prime\rangle &=& \frac{ e^{-H^\prime} + e^{-2K^\prime}}{e^{H^\prime} +
 e^{-H^\prime} + 2e^{-2K^\prime}} =
 \frac{1}{2} \left( 1 - \frac{ e^{H^\prime} - e^{-H^\prime}}{e^{H^\prime} +
 e^{-H^\prime} + 2e^{-2K^\prime}} \right)
\end{eqnarray}
It is easy to see that these general expressions reduce to the
results of Eqs~\eqref{diss} and \eqref{hall} in the limit of low
temperatures and small $H$.
\subsection{Chiral fermions \label{chiralF1}}
We next introduce the idea of finite system sizes in the theory of
chiral fermions. For this purpose we write the action of Eq.
\eqref{SCG0} as an integral over the finite interval $0\leqslant
x\leqslant L$
\begin{equation}
S_\textrm{eff} = \int_0^{L} dx  \bar{\Psi} (i\partial_x + i
H\tau_z + i e^{-2K}\tau_x)\Psi .\label{SeffCF}
\end{equation}
Assuming anti-periodic boundary conditions on the fermion fields
for simplicity then we can express the derivative $i\partial_x$ in
terms of a discrete set of frequencies $\omega_n = L^{-1} \pi
(2n+1)$. Next, by interpreting $x$ as the imaginary time and $L$
as the inverse temperature then the action of Eq.~\eqref{SeffCF}
corresponds to the following spin hamiltonian
\begin{equation}
\mathcal{H} = - H\tau_z - e^{-2K}\tau_x\label{HamCF} .
\end{equation}
From the analysis on boundary conditions in the previous Sections
we infer that Eq. \eqref{free} can be expressed in the hamiltonian
formalism according to
\begin{equation}
 Z[q, \beta a_0] =  \frac{ \Tr  B (q, \beta a_0 )
 e^{-L\mathcal{H}}}{\Tr B (0,0)
 e^{-L\mathcal{H}}}.\label{freefermions}
\end{equation}
The equivalence of Eqs \eqref{freefermions} and \eqref{free1aa} is
readily established once it is recognized that in the limit of
small $e^{-2K}$ and $H$ we can write
\begin{equation}
 T^L = e^{2LK} e^{-L\mathcal{H}} .
\end{equation}
The results obtained from chiral fermion theory are therefore
identically the same as those obtained from the Ising model, Eq.
\eqref{free1a}, with the conductances given as in Section
\ref{ising2}.
\subsection{Coulomb gas \label{explicit}}
\subsubsection{Conductances}
As an important check on the results derived in Sections
\ref{ising2} and \ref{chiralF1} we next present the expressions
for the conductance parameters $\langle\sigma_{xx}^\prime\rangle$
and $\langle\sigma_{xy}^\prime\rangle$ as obtained directly from
the Coulomb gas representation in a computation to second order in
the fugacity $\sigma/\beta$. To start, we first list the results
for the partition function which can be written as in Eq.
\eqref{partition}, i.e.
\begin{equation}
Z[q,\beta a_0,\theta] = \sum_{m\in \mathbb{Z}} \zeta(m)
e^{-\frac{L}{4\kappa}(m+\theta/2\pi)^2 -i 2\pi m  q} .\label{Znew}
\end{equation}
The complete expression for $\zeta(m)$ to second order in
$\sigma/\beta$ is as follows
\begin{eqnarray}
 \zeta(m) = 1 &-& \frac{8 \kappa\sigma}{\beta}  \cos
 \beta a_0 \left[ \frac{e^{i \pi q}}{2m -1+\frac{\theta}{\pi}} -
 \frac{e^{-i \pi q}}{2m +1+\frac{\theta}{\pi}} \right] \\
 &+& \left( \frac{4\kappa \sigma}{\beta} \right)^2 \cos 2\beta a_0
 \Bigl [ -\frac{2}{(2m+1+\frac{\theta}{\pi})(2m-1+\frac{\theta}{\pi})}\notag \\
 &&\hspace{4cm}
 +\frac{e^{i 2\pi q}}{(2m-2+\frac{\theta}{\pi})(2m-1+\frac{\theta}{\pi})}\notag\\
 &&\hspace{4cm}+ \frac{e^{-i2\pi q}}{(2m+2+\frac{\theta}{\pi})
 (2m+1+\frac{\theta}{\pi})}\Bigr ]
 \notag\\
 &+& \left( \frac{4\kappa \sigma}{\beta} \right)^2
 \Bigl [ \frac{L}{4\kappa}
 \left( \frac{1}{2m+1+\frac{\theta}{\pi}}-\frac{1}{2m-1+\frac{\theta}{\pi}} \right)
 \notag \\
 &&\hspace{2cm}- \frac{1-e^{i2\pi q}}{(2m-1+\frac{\theta}{\pi})^2}
 -\frac{1-e^{-i 2\pi q}}{(2m+1+\frac{\theta}{\pi})^2} \Bigr ] .
 \label{q3}
\end{eqnarray}
The contributions proportional to $\cos2\beta a_0$ arise from the
terms with charges $n_+=0,2$ and $n_-=2,0$ in the interior of the
system, see Eq. \eqref{Zmm}. Similarly, the other contributions of
the order $(\sigma/\beta)^2$ originate from the terms with
$n_+=n_-=1$. Notice that Eq.~\eqref{Znew} has the following
symmetry
\begin{equation}
 Z[q,\beta a_0,2\pi-\theta] = e^{i\,2\pi q} Z[-q,\beta a_0,\theta] .
\end{equation}
This result ensures that the ``particle-hole symmetry" is retained
by both the free energy and the conductances (see
Eq.~\eqref{conduct})
\begin{eqnarray}
 F(2\pi-\theta)&=&F(\theta) \\
 \langle \sigma_{xx}^\prime (2\pi-\theta) \rangle &=&
 \langle \sigma_{xx}^\prime (\theta) \rangle \\
 \langle \sigma_{xy}^\prime (2\pi-\theta) \rangle &=&
 1- \langle \sigma_{xy}^\prime (\theta) \rangle .
\end{eqnarray}
Next, considering the limit $\theta\to\pi^{-}$ then the sum in
Eq.~\eqref{Znew} is dominated by the terms with $m=0,-1$ only. The
results for the conductances can be written as follows
\begin{eqnarray}
 \langle \sigma_{xx}^\prime \rangle &=&
 \frac{\eta_+(\omega)}{f(\omega) e^{\omega L}+f(-\omega)e^{-\omega L}
 +2\eta_+(\omega)} \label{1a}
 \\
 \langle \sigma_{xy}^\prime \rangle &=& \frac{1}{2}\left [1-
 \frac{f(\omega) e^{\omega L}-f(-\omega)e^{-\omega L} + 2\eta_-(\omega)}
 {f(\omega) e^{\omega L}+f(-\omega)e^{-\omega L}+2\eta_+(\omega)} \right ]
 \label{1b}
\end{eqnarray}
with the following meaning of the symbols
\begin{eqnarray}
 \eta_{\pm} & = & g_{\pm}(\omega)e^{\omega L}\pm
 g_{\pm}(-\omega)e^{-\omega L} \\
 \nonumber \\
 f(\omega) &=& 1+ \left( \frac{\sigma}{\beta\omega}\right)^2 \frac{
 L\omega}{1-4\kappa\omega} \left [1-\frac{6\kappa}{L}\left
 (1+\frac{32\kappa\omega(1-4\kappa\omega)}
 {(1+8\kappa\omega)(3-8\kappa\omega)}\right )\right ] \label{fomega} \\
 g_+(\omega) &=& \left( \frac{\sigma}{\beta\omega}\right)
 \frac{1}{2(1-4\kappa\omega)}+\left(
 \frac{\sigma}{\beta\omega}\right)^2
 \frac{2 (8\kappa\omega)^2}{(1+8\kappa\omega)(3-8\kappa\omega)} \\
 g_{-}(\omega)&=& \left( \frac{\sigma}{\beta\omega}\right)
 \frac{4\kappa\omega}{1-4\kappa\omega}- \frac{1}{4} \left(
 \frac{\sigma}{\beta\omega}\right)^2
 \Bigl [ 1+ \frac{8\kappa\omega}{1+8\kappa\omega}  -
 \frac{(4\kappa\omega)^2}{(1-4\kappa\omega)^2} \notag \\
 &&\hspace{2cm}-
 \frac{2(4\kappa\omega)^2}{(1-4\kappa\omega)(3-8\kappa\omega)}-
 \frac{8(8\kappa\omega)^2}{(1+8\kappa\omega)(3-8\kappa\omega)}
 \Bigr ].\label{gminus}
\end{eqnarray}
Notice that the functions $f$ and $g_\pm$ are given in terms of a
regular series expansion in powers of $\sigma/(\beta\omega)$,
$\omega L$ as well as $\kappa\omega$. To understand these results
let us first write Eqs \eqref{1a} and \eqref{1b} in a slightly
more transparent manner according to
\begin{eqnarray}
\langle \sigma_{xx}^\prime \rangle &=&
\frac{\tilde{\eta}_+(\omega)}{ e^{X}+
e^{-X}+2\tilde{\eta}_+(\omega)} \label{2a}
\\
\langle \sigma_{xy}^\prime \rangle &=& \frac{1}{2}\left [1- \frac{
e^{X}- e^{-X}+2\tilde{\eta}_-(\omega)}{ e^{X}+
e^{-X}+2\tilde{\eta}_+(\omega)} \right ] \label{2b}
\end{eqnarray}
where
\begin{eqnarray}
 X & = & L \omega +\frac{1}{2} \ln \frac{f(\omega)}{f(-\omega)} \\
 \tilde{\eta}_\pm &=& \frac{{\eta}_\pm}{\sqrt{f(\omega)
 f(-\omega)}}\label{Xff}
\end{eqnarray}
Next, it is not difficult to see that Eqs \eqref{2a} - \eqref{Xff}
precisely correspond to the results listed in Section \ref{ising2}
provided we drop all the terms with $\kappa\omega$ and $\kappa/L$
in the expressions for $f$ and $g_\pm$. The correct way of
expressing this is by saying that we are interested in the
thermodynamic limit $L \approx \beta \rightarrow \infty$ while
keeping the quantities $\omega L$ and $\sigma/(\beta\omega)$
fixed. Under these circumstances we have $\kappa\omega =
\mathcal{O}(L^{-2})$ and $\kappa/L = \mathcal{O}(L^{-2})$ both of
which therefore vanish. Keeping only the surviving terms we can
write
\begin{eqnarray}
 X & \rightarrow & \omega L \left (1+
 \frac{1}{2}\left (\frac{\sigma}{\beta\omega}\right )^2\right) \\
 \tilde{\eta}_+ & \rightarrow & \eta = \frac{\sigma}{\beta \omega}\sinh \omega L \\
 \tilde{\eta}_- & \rightarrow & ~- \frac{\sigma^2}{\beta^2\omega^2}\sinh\omega L
\end{eqnarray}
These results can be written precisely in the form of Eqs
\eqref{diss} and \eqref{hall} with $X$ given in terms of a series
expansion in powers of $\sigma/(\beta\omega)$
\begin{equation}
 X = L \sqrt{\omega^2 + \left( \frac{\sigma}{\beta} \right)^2}
 = \omega L \left( 1 +
 \frac{1}{2} \left( \frac{\sigma}{\beta \omega} \right)^2  +\dots
 \right).
\end{equation}
Similarly, the square root in Eq. \eqref{hall} is expanded
according to
\begin{equation}
 \sqrt{(e^X - e^{-X})^2 -4\eta^2} = (e^X -
 e^{-X})\left( 1 - \frac{1}{2} \left( \frac{\sigma}{\beta\omega}\right)^2
 + \dots \right) .\label{squareroot}
\end{equation}
In summary, the expressions for the conductances as derived in
Sections \ref{ising2} and \ref{chiralF1} are entirely consistent
with those obtained from the Coulomb gas representation to order
$(\sigma/\beta)^2$. One may in principle proceed and introduce
several different renormalization constants $Z$ that absorb all or
parts of the corrections of order $\kappa\omega$ and $\kappa/L$ in
Eqs \eqref{fomega} -\eqref{gminus}. For instance, the complete
expression for $X$ in Eq. \eqref{Xff} can be written in the form
\begin{equation}
 X = L \sqrt{Z_\omega^2 \omega^2 + \left( \frac{\sigma}{\beta} \right)^2}
\end{equation}
where to the appropriate order in $\sigma/\beta$ the coefficient
$Z_\omega$ is equal to
\begin{equation}
 Z_\omega = 1+ \left( \frac{4\kappa\sigma}{\beta}\right)^2
 \left [ \frac{1-{6\kappa}/{L}}{1-(4\kappa\omega)^2} +
 \frac{ {96\kappa}/{L}  }
 {(1-(8\kappa\omega)^2)(9-(8\kappa\omega)^2)}\right
 ] .
\end{equation}
Similarly, one can discuss the corrections in Eq.
\eqref{squareroot} but the expressions are somewhat cumbersome and
generally different from those entering the correlation functions.
\subsubsection{Distribution of quantum Hall states}
For completeness we next present the final total result for the
partition function of Eq. ~\eqref{Znew} which is generally more
complex than the expressions encountered sofar. Eq. ~\eqref{Znew}
can most conveniently be written as follows
\begin{equation}
Z[q,\beta a_0,\theta] = e^{-F(\theta)} \sum_{j=-2}^4
\mathcal{K}_j(\beta a_0) e^{ij\pi q} \label{sumj}
\end{equation}
where
\begin{eqnarray}
 \mathcal{K}_{-2} &=& p_{-2,0}+ p_{-2,2} \cos 2 \beta a_0 \notag \\
 \mathcal{K}_{-1} &=& ~~~~~~~~~~ p_{-1,1}\cos \beta a_0 \notag \\
 \mathcal{K}_0 &=& \mathcal{P}_0 ~~~~+p_{0,2}\cos 2 \beta a_0 \notag \\
 \mathcal{K}_1 &=& ~~~~~~~~~~~~~ \mathcal{P}_\pi \cos \beta a_0\notag \\
 \mathcal{K}_2 &=& \mathcal{P}_{2\pi} + p_{2,2} \cos 2 \beta a_0 \notag \\
 \mathcal{K}_3 &=& ~~~~~~~~~~~~ p_{3,1}\cos \beta a_0 \notag \\
 \mathcal{K}_4 &=& p_{4,0}+p_{4,2} \cos 2 \beta a_0. \label{probPAR}
\end{eqnarray}
Here, the quantities $\mathcal{K}_j (\beta a_0)$ obey the general
constraints
\begin{equation}
\sum_{j=-2}^4\mathcal{K}_j(0)=1,\,\,\sum_{j=-2}^4
j\mathcal{K}_j(0)=2\langle \sigma_{xy}^\prime \rangle,\,\,\,
\sum_{j=-2}^4 \mathcal{K}_j^{\prime\prime}(0)=2\langle
\sigma_{xx}^\prime \rangle .\label{ConstPAR}
\end{equation}
Based on these constraints one can express the $\mathcal{P}_0$,
$\mathcal{P}_\pi$ and $\mathcal{P}_{2\pi}$ in Eqs \eqref{probPAR}
in terms of the quantities $p_{i,j}$ as well as the conductances
$\langle \sigma_{xy}^\prime \rangle$ and $\langle \sigma_{x
x}^\prime \rangle$ as given by Eqs \eqref{1a} and \eqref{1b}. The
result is
\begin{eqnarray}
\mathcal{P}_0 &=& 1-\langle \sigma_{xy}^\prime \rangle-\langle
 \sigma_{xx}^\prime \rangle + \mathcal{P}^\prime_0 \notag \\
 \mathcal{P}_\pi &=& ~~~~~~~~~~~~~~ 2\langle
 \sigma_{xx}^\prime \rangle + \mathcal{P}^\prime_\pi \notag \\
 \mathcal{P}_{2\pi} &=& ~~~~~ \langle \sigma_{xy}^\prime \rangle-\langle
 \sigma_{xx}^\prime \rangle + \mathcal{P}^\prime_{2\pi}\label{Ps}
\end{eqnarray}
where
\begin{eqnarray}
\mathcal{P}^\prime_0 &=&  -2p_{4,0} + p_{-2,0} ~+~p_{2,2}
+2p_{0,2} -p_{3,1}
 +p_{-1,1} ~~~~~~~~~~ +3p_{-2,2} \notag\\
 \mathcal{P}^\prime_\pi &=& ~~~~~~~~~~~~~~~~~~~ -4p_{2,2} -4p_{0,2} -p_{3,1}
 -p_{-1,1} -4p_{4,2} -4p_{-2,2} \notag\\
 \mathcal{P}_{2\pi}^\prime &=& \hspace{0.53cm}p_{4,0} -2 p_{-2,0} +2p_{2,2}~ +~p_{0,2} +p_{3,1}
 -p_{-1,1} +3p_{4,2} .
\end{eqnarray}
The quantities $p_{j,k}$ in Eqs \eqref{probPAR} indicate that the
transition between the $\theta=0$ and $\theta=2\pi$ vacuum
generally involves a range of different $\theta$ vacua or quantum
Hall states. In total 8 different quantities $p_{j,k}$ are given
by
\begin{eqnarray}
 p_{-2,0} &=& \left (\frac{2\kappa\sigma}{\beta}\right )^2
 \frac{e^{\omega L}}{(1-4\kappa\omega)^2}
 D^{-1}  \notag  \\
 p_{-2,2}&=& 2\left (\frac{2\kappa\sigma}{\beta}\right )^2
 \frac{e^{\omega L}}{(1-4\kappa\omega)(3-8\kappa\omega)} D^{-1}
 \notag \\
 p_{-1,1}&=& \left (\frac{4\kappa\sigma}{\beta}\right )
 ~\frac{e^{\omega L}}{1-4\kappa\omega} D^{-1}
 \notag\\
 p_{0,2} &=& \left (\frac{2\kappa\sigma}{\beta}\right )
 \left (\frac{\sigma}{\beta\omega}\right )
 \left( \frac{e^{\omega
 L}}{1+4\kappa\omega}- \frac{e^{-\omega L}}{1-8\kappa\omega}
 \right) D^{-1} \notag  \\
 p_{2,2}&=& \left (\frac{2\kappa\sigma}{\beta}\right )
 \left (\frac{\sigma}{\beta\omega}\right ) \left( \frac{e^{\omega
 L}}{1+8\kappa\omega}- \frac{e^{-\omega L}}{1-4\kappa\omega}
 \right) D^{-1} \notag  \\
 p_{3,1}&=&\left (\frac{4\kappa\sigma}{\beta}\right )
 ~\frac{e^{-\omega
 L}}{1+4\kappa\omega} D^{-1} \notag\\
p_{4,0} &=& \left(\frac{2\kappa\sigma}{\beta}\right)^2
 \frac{e^{-\omega L}}{(1+4\kappa\omega)^{2}} D^{-1}
 \notag \\
p_{4,2} &=& 2\left (\frac{2\kappa\sigma}{\beta}\right )^2
 \frac{e^{-\omega L}}{(1+4\kappa\omega)(3+8\kappa\omega)} D^{-1}
 \label{pPAR}
\end{eqnarray}
where
\begin{eqnarray}
 D={f(\omega) e^{\omega L}+f(-\omega)e^{-\omega L}
 +2\eta_+(\omega)} .
\end{eqnarray}
To lowest order in an series expansion in powers of $\sigma/\beta$
and $\kappa\omega$ we can write these results as follows
\begin{eqnarray}
 p_{-2,0} &\approx& ~~~~\left
 (\frac{2\kappa\sigma}{\beta}\right )^2 ~
 (1-\langle \sigma_{xy}^\prime \rangle) \notag  \\
 p_{-2,2}&\approx& ~~\frac{2}{3} \left (\frac{2\kappa\sigma}{\beta}\right )^2
 ~(1-\langle \sigma_{xy}^\prime \rangle) \notag \\
 p_{-1,1}&\approx& ~~~~ \left
 (\frac{4\kappa\sigma}{\beta}\right )~~
 ~(1-\langle \sigma_{xy}^\prime \rangle) \notag\\
 p_{0,2} &\approx&  -\frac{3}{2}
 \left (\frac{4\kappa\sigma}{\beta}\right )^2
 ~(1-\langle \sigma_{xy}^\prime \rangle)~~+
 \left (\frac{4\kappa\sigma}{\beta}\right )
 \langle \sigma_{xx}^\prime \rangle \notag
\\
 p_{2,2}&\approx&  -\frac{3}{2}
 \left (\frac{4\kappa\sigma}{\beta}\right )^2
 ~\langle \sigma_{xy}^\prime \rangle ~~~~~~~~~~+
 \left (\frac{4\kappa\sigma}{\beta}\right )
 \langle \sigma_{xx}^\prime \rangle\notag  \\
 p_{3,1} &\approx& ~~~~\left
 (\frac{4\kappa\sigma}{\beta}\right )
 ~~\langle \sigma_{xy}^\prime \rangle \notag \\
 p_{4,0} &\approx& ~~~~\left (\frac{2\kappa\sigma}{\beta}\right )^2
 ~\langle \sigma_{xy}^\prime \rangle \notag
\end{eqnarray}
\begin{eqnarray}
 p_{4,2} &\approx& ~~\frac{2}{3} \left
 (\frac{2\kappa\sigma}{\beta}\right )^2
 ~\langle \sigma_{xy}^\prime \rangle \notag \\
 \label{pPAR1}
\end{eqnarray}
The factors $\kappa\sigma/\beta$ indicate that the quantities
$p_{i,j}$ are exponentially small corrections terms that can be
ignored relative to the leading order terms contained in
$\mathcal{P}_0$, $\mathcal{P}_\pi$ and $\mathcal{P}_{2\pi}$, Eqs
\eqref{Ps}. Perhaps the most important conclusion that one can
draw from the results of this Section is that both the {\em
robustly} quantized quantum Hall plateau and the quantum critical
behavior of the plateau {\em transitions} simultaneously emerge
from the existence of {\em discrete topological sectors} in the
theory. These topological sectors, labelled by the integer $j$ in
Eq. \eqref{sumj}, have not been recognized in the historical
papers on the large $N$ expansion. They are nevertheless one of
the most fundamental features of the instanton vacuum concept in
scale invariant theories.
\section{\label{RGLN} Scaling diagram for the large N
expansion}
Discarding the corrections to scaling discussed in the previous
Section we next return to the expressions for the conductances
$\langle\sigma_{xx}^\prime\rangle $ and
$\langle\sigma_{xy}^\prime\rangle $, Eqs \eqref{diss} -
\eqref{etaweak}, which are some of the most important results of
this paper. We have already mentioned several times earlier that
these expressions have an entirely different significance
depending on the physical context in which they are being used.
These differences are clearly reflected in the renormalization
behavior of the theory which is the main topic of the present
Section.
\subsection{Renormalization. Ising model}
Let us first discuss the $1D$ Ising model which is in many ways
standard. Eqs~\eqref{diss} - \eqref{etaweak} involve two different
scaling variables, namely
\begin{equation}
 H^\prime =LH  ,
 ~~ e^{-2K^\prime} = L e^{-2K} .\label{substising}
\end{equation}
Here, $H^\prime$ and $K^\prime$ are the low temperature and small
$H$ versions of the more general expressions given by Eqs
\eqref{Kprime} and \eqref{Hprime}. The renormalization group
equations for small values of $H$ can generally be expressed as a
series expansion in powers of $K^{-1}$
\begin{equation}
 \frac{d H^\prime}{d\ln L} = H^\prime \left( 1 + \mathcal{O}(1/K^\prime) \right),
 \qquad
 \frac{d K^\prime}{d\ln L} = -\frac{1}{2}  + \mathcal{O}(1/K^\prime).
 \label{betaising}
\end{equation}
These results show that the $1D$ Ising model is a prototypical
example of an asymptotically free field theory with interesting
features such dynamic {\em mass generation}.
\subsection{Renormalization. Coulomb gas}
Next we turn to the Coulomb gas problem which has a different
dimensionality. Substituting the parameters of the large $N$
expansion for the Ising model variables $H$ and $e^{-2K}$ we
obtain
\begin{equation}
 LH = L\omega = \left (1-\frac{\theta}{\pi}\right )
 \frac{6\pi M^2 L\beta}{N} ,
 \qquad L e^{-2K} = L \frac{\sigma}{\beta}
 = N M L \frac{e^{-M\beta}}{\sqrt{2\pi M\beta}} .\label{subst}
\end{equation}
To discuss the role of the parameter $N$ we introduce an arbitrary
scale factor $b$ according to
\begin{equation}
 b^2 = \frac{M^2 L\beta}{N} \gg 1 .
\end{equation}
Equation \eqref{subst} can then be written as
\begin{equation}
 LH = 6\pi b^2 \left (1-\frac{\theta}{\pi}\right ),
 \qquad L e^{-2K} =
 {\alpha^2 n^2} \sqrt{\frac{nb}{2\pi}} e^{-nb} \label{subst1}
\end{equation}
where $\alpha$ and $n$ are defined as follows
\begin{equation}
 \alpha = L/\beta ,\qquad  n=\sqrt{N/\alpha}. \label{subst2}
\end{equation}
Since the quantities $b$ and $n$ in Eq. \eqref{subst1} are
independent variables we conclude that the large $N$ limit of the
theory is generally well defined and obtained by taking $n
\rightarrow \infty$ first while keeping scale factor $b$ fixed.
This definition furthermore ensures that the large $N$ expansion
and the strong coupling expansion of the Coulomb gas in powers of
the fugacity are mutually consistent. Next we combine Eqs
\eqref{subst1} and \eqref{substising} and express the
renormalization of the Coulomb gas in terms of the Ising model
quantities $H^\prime$ and $K^\prime$. Keeping in mind that we now
have $K^\prime = \mathcal{O}(n)$ then the renormalization group
equations for large values of $n$ are obtained as follows
\begin{equation}
 \frac{d H^\prime}{d\ln b} = 2 H^\prime ,
 \qquad
 \frac{d K^\prime}{d\ln b} = K^\prime \left[1 +
 \mathcal{O}\left (\frac{\ln K^\prime}{K^\prime}\right ) \right ].
 \label{betalargen}
\end{equation}
Unlike the Ising model, the results now indicate that the theory
along the line $H^\prime = 0$ or $\theta=\pi$ generally displays
{\em gapless} excitations, rather than a {\em mass gap}. Moreover,
the result for $H^\prime$ is in accordance with the fact that the
transition at $\theta=\pi$ is a first order one and, at the same
time, displays a divergent correlation length with an exponent
$1/2$.

\begin{figure}[tbp]
\begin{center}
\includegraphics[width=120mm]{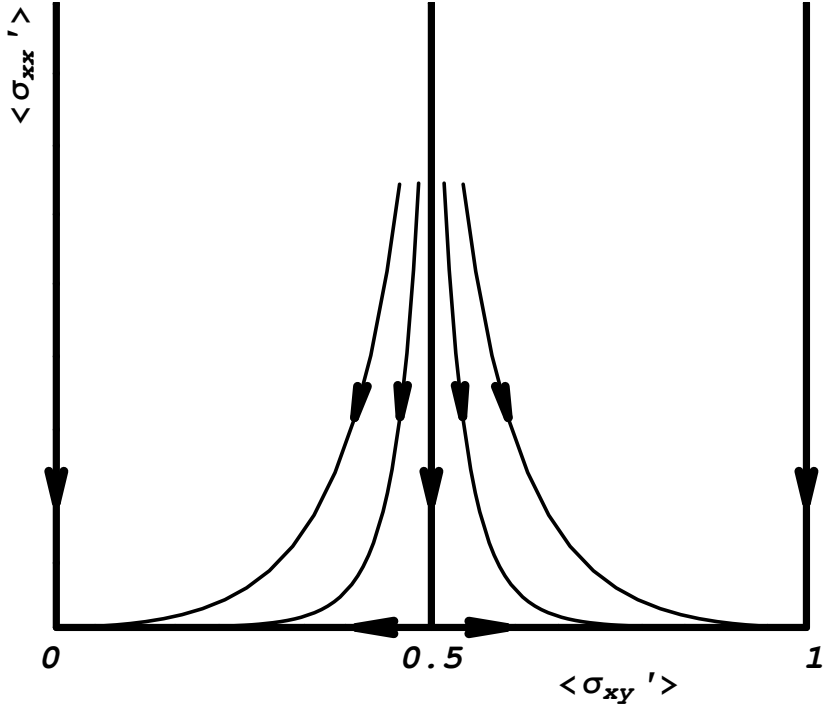}
\caption{Renormalization group flow in the $\langle
\sigma^\prime_{xx}\rangle$ and $\langle \sigma^\prime_{xy}\rangle$
conductance plane according to Eqs \eqref{diss} - \eqref{etaweak}
and \eqref{subst1}, see text.}\label{Fig1}
\end{center}
\end{figure}

Since the leading order results of Eq. \eqref{betalargen} do not
contain the parameter $N$ or $n$ explicitly, we can make use of
Eqs \eqref{substising} and \eqref{diss} - \eqref{etaweak} and
project the renormalization group flow lines of the large $N$
theory directly onto the $\langle\sigma_{xx}^\prime\rangle $,
$\langle\sigma_{xy}^\prime\rangle $ conductance plane. For
illustration we have plotted in Fig. \ref{Fig1} the results
obtained from numerical simulation. It is interesting to notice
that the flow diagrams of Figs \ref{RGFig1s} and \ref{Fig1} are
numerically very similar. This is in spite of the fact that the
previously obtained scaling results (Eqs \eqref{kubo1}
-\eqref{XetaCG2}) and those of the present paper (Eqs \eqref{diss}
- \eqref{etaweak}) are qualitatively very different.

%
\subsection{Regularity condition}
As a final step in this paper we next make sure that the square
root singularities appearing in Eqs \eqref{diss} - \eqref{etaweak}
do not spoil the regularity conditions of the renormalization
group equations. To investigate this point further we compute the
$\beta$ functions
\begin{eqnarray}
 \frac{d \langle\sigma_{xx}^\prime\rangle}{d\ln b} =
 \beta_{xx}(\langle\sigma_{xx}^\prime\rangle,
 \langle\sigma_{xy}^\prime\rangle ) \\
 \frac{d \langle\sigma_{xy}^\prime\rangle}{d\ln b}
 =\beta_{xy}(\langle\sigma_{xx}^\prime\rangle,
 \langle\sigma_{xy}^\prime\rangle )
\end{eqnarray}
based on the leading order results of Eq. \eqref{betalargen}. We
obtain the following expressions
\begin{eqnarray}
\beta_{xx}(k,h) &=& k \left \{1-
 2k\frac{(1-2h)^2+4k [f(k,1-2h)-1+2k] }{(1-2h)^2+4k^2}\right \}
 \notag \\
 &\times & \ln\frac{4k
 f(k,1-2h)}{1-4k-(1-2h)^2}
 + k \frac{2 (1-2h)^2
 [f(k,1-2h)-1+2k]}{(1-2h)^2+4k^2}\notag \\
 &&\label{RG1} \\
 \beta_{xy}(k,h) &=& \left(2h-1\right )\left [ f(k,1-2h) +k\right ]
 -k \left(2h-1\right )\ln\frac{4e^{-1} k f(k,1-2h)}{1-4k-(1-2h)^2}\notag \\
 &\times & \left \{1-\frac{4k\left
 [f(k,1-2h)-1+2k\right ]}{(1-2h)^2+4k^2} \right \}.
 \label{RG2}
\end{eqnarray}
Here we have introduced the function
\begin{equation}
 f(k,h) = \frac{1-4k -h^2}{2\sqrt{h^2+4k^2}}\ln
 \frac{1-2k+\sqrt{h^2+4k^2}}{1-2k-\sqrt{h^2+4k^2}}.
\end{equation}
To study the behavior of $\beta_{xx}$ and $\beta_{xy}$ in the
fixed point regime $k\ll 1$ and $|1-2h|\ll 1$ we quote the
following results which are valid for arbitrary values of
$|1-2h|/k$
\begin{eqnarray}
\beta_{xx} (k,h) &\approx& k \left [1-\frac{2k
(1-2h)^2}{(1-2h)^2+4k^2}\right ]\ln 4k -\frac{4 k
(1-2h)^2}{3}\label{RG2_1} \\
\beta_{xy} (k,h) &\approx& (2h-1)\left
(1-\frac{2(1-2h)^2}{3}-k \ln 4k \right ). \label{RG1_2}
\end{eqnarray}
In the limit where $k$ goes to zero and $h$ approaches $1/2$ we
therefore obtain
\begin{equation}
 \beta_{xx} (k,1/2) = k \ln k ,\qquad
 \beta_{xy} (0,h) = 2h-1 . \label{RG2_3}
\end{equation}
Notice that these results are consistent with those of Eq.
\eqref{betalargen} under the identification $k=e^{-2K^\prime}$ and
$h= \frac{1}{2} - H^\prime$.
\section{Conclusion \label{Conclusion}}
Starting from the Coulomb gas representation of the $CP^{N-1}$
model with large values of $N$ in two dimensions we have
identified the exact critical theory for the transition at
$\theta=\pi$. This theory is one dimensional and none other than
the theory of massless chiral fermions that has previously emerged
in the theory of the quantum Hall effect, in particular the
Luttinger liquid theory of edge excitations~\cite{MasslessNew}.

We have benefitted from the various alternative approaches that we
have introduced, in particular the hamiltonian approach as well as
the mapping of the Coulomb gas onto the $1D$ Ising model. Besides
computational advantages, these different mappings also elucidate
the role played by the {\em geometry} of the system in general,
and the meaning of topics such as {\em mass generation} at $\theta
= \pi$ in particular.

Perhaps the most interesting conclusion of this paper is that the
divergent correlation length $\xi \propto |\theta - \pi|^{-1/2}$
emerges not only from the physical objectives of the quantum Hall
effect, but also from Coleman's original ideas on the transition
at $\theta=\pi$~{\cite{Coleman}}. Based on an explicit knowledge
of the multi-point correlation functions we have shown that the
mechanism responsible for changing the total number of charged
particles at the edges of the system is in fact synonymous for the
existence of {\em gapless} bulk excitations at $\theta=\pi$.
Remarkably, neither the existence of these excitations nor the
significance of Coleman's mechanism in terms of quantum Hall
physics has previously been
recognized~{\cite{Affleck85,Affleck88}}.

The results of this paper provide the complete conceptual
structure that one in general can associate with the topological
concept of an instanton vacuum. Besides an exactly solvable {\em
critical theory} for $\theta \approx \pi$ this structure
furthermore includes finite size {\em scaling results} for the
``conductances", {\em robust} topological quantum numbers that
explain the precision and stability of the quantum Hall plateaus
and also the {\em massless chiral edge excitations} that
facilitate the flow of Hall currents. The fundamental features of
the quantum Hall effect are therefore not merely a topic of
replica field theory or disordered free electron systems alone.
Rather than that, they are a {\em super universal} phenomenon that
teaches us something fundamental about instanton vacuum in
asymptotically free field theory in general.

As a final remark, it should be mentioned that the statement of
{\em super universality} has recently been investigated and
studied, with great success, in several completely different
physical systems such as the theory in the presence of
electron-electron interactions,~\cite{PruiskenBurmistrov2} quantum
spin chains~\cite{PruiskenShankarSurendran} as well as the
Ambegaokar-Eckern-Sch\"on theory of the Coulomb blockade
problem~\cite{PruiskenBurmistrov3}.

\section{Acknowledgments}

This research was initiated during the Amsterdam Summer Workshop
``Low-D Quantum Condensed Matter''. The authors are grateful to
the participants, in particular A.\,G.\,Abanov, for discussions.
The work was funded in part by the Dutch National Science
Foundations \textit{NWO} and \textit{FOM}. One of us
(\textit{ISB}) is indebted to P.\,M.\,Ostrovsky for fruitful
discussions and to Russian Science Support Foundation, the Russian
Ministry of Education and Science and Council for Grants of the
President of Russian Federation for financial support.

\appendix
\setcounter{addeq}{\value{equation}} \setcounter{equation}{0}
\renewcommand{\theequation}{A.\arabic{equation}}
\section{\label{AppA} Two-point correlation function $g_2$ in the Ising model}

In this appendix we present a brief derivation of the
result~\eqref{adaga} in the Ising model representation. As
well-known, the standard approach to exact solution the Ising
model is via transfer matrix technique. For $1D$ Ising model the
transfer matrix is given by Eq.~\eqref{TransfMat}. It can be
diagonalized by an orthogonal rotation
\begin{equation}
U = e^{i \phi \tau_y}, \qquad \phi = \frac{1}{2} \arcsin
\frac{e^{-2K}}{\sqrt{\sinh^2 H + e^{-4K}}}.
\end{equation}
The eigenvalues of the $T$ are
\begin{equation}
\lambda_\pm = e^K \left (\cosh H \pm \sqrt{\sinh^2 H + e^{-4K}}
\right ).
\end{equation}
The domain wall operators $a_{+-}$ and $a_{-+}$ can be expressed
in the space in which the transfer matrix acts as follows
\begin{equation}
a_{+-} = \frac{\tau_x+i\tau_y}{2},\qquad a_{-+} =
\frac{\tau_x-i\tau_y}{2}.
\end{equation}
Defining the two-point correlation function $g_2(i,j)$ as
\begin{equation}
g_2(i,j)=\langle a_{+-}(i) a_{-+}(j) \rangle = \lim \limits_{M\to
\infty} \frac{\tr T^i a_{+-} T^{j-i} a_{-+} T^{M-j}}{\tr T^M}
\end{equation}
we easily find
\begin{equation}
g_2(i,j) = m_0 + m_+ \vartheta(j-i) e^{-|j-i|/\xi}+ m_-
\vartheta(i-j) e^{-|j-i|/\xi}.\label{adagaIsing}
\end{equation}
Here the coefficients are given by Eqs~\eqref{m0}-\eqref{m+} with
the magnetization $\mathcal{M}$ defined in
Eq.~\eqref{MagnetIsing}. The result~\eqref{adagaIsing} involves
the correlation length $\xi$ of $1D$ Ising model
\begin{equation}
\xi = \frac{1}{\ln\lambda_+/\lambda_-} = \tanh^{-1} \frac{\tanh
H}{\mathcal{M}}.
\end{equation}
In the limit of weak magnetic field $H\ll 1$ the
result~\eqref{adagaIsing} coincides with the result~\eqref{adaga}
for $g_2$ obtained in the framework of $1D$ chiral fermion theory.


\end{document}